\documentclass[printer]{aa}

\usepackage{txfonts}
\usepackage{graphicx}
\usepackage{array}
\usepackage{booktabs}
\usepackage{amssymb}
\usepackage{subcaption}
\usepackage{multirow}
\usepackage{url}
\usepackage[colorlinks=true,allcolors=blue]{hyperref}
\PassOptionsToPackage{hyphens}{url}
\usepackage{xcolor}

\def\Msun{M_\odot}

\def\Teff{\ensuremath{T_\mathrm{eff}}\xspace}

\newcommand{\ffarcs}{\mbox{\ensuremath{.\!\!^{\prime\prime}}}}

\newcolumntype{L}[1]{>{\raggedright\let\newline\\\arraybackslash\hspace{0pt}}m{#1}}
\newcolumntype{C}[1]{>{\centering\let\newline\\\arraybackslash\hspace{0pt}}m{#1}}

\begin{document}

\title{Searching for low-mass companions at small separations in transition disks with aperture masking interferometry\thanks{Based on observations collected at the European Southern Observatory under ESO programmes 60.A-9800(S), 198.C-0209(G), 1100.C-0481(F), 1100.C-0481(N), 105.2067.001}}

\titlerunning{Searching for low-mass companions in transition disks with aperture masking interferometry}

\author{
T.\,Stolker \inst{\ref{leiden}}
\and J.\,Kammerer\inst{\ref{stsci}}
\and M.\,Benisty\inst{\ref{oca},\ref{grenoble}}
\and D.\,Blakely\inst{\ref{victoria},\ref{herzberg}}
\and D.\,Johnstone\inst{\ref{victoria},\ref{herzberg}}
\and M.\,L.\,Sitko\inst{\ref{boulder}}
\and J.\,P.\,Berger\inst{\ref{grenoble}}
\and J.\,Sanchez-Bermudez\inst{\ref{unam}, \ref{mpia}}
\and A.\,Garufi\inst{\ref{inaf}}
\and S.\,Lacour\inst{\ref{lesia},\ref{eso}}
\and F.\,Cantalloube\inst{\ref{marseille}}
\and G.\,Chauvin\inst{\ref{oca}}
}

\institute{
Leiden Observatory, Leiden University, Niels Bohrweg 2, 2333 CA Leiden, The Netherlands\\
\email{stolker@strw.leidenuniv.nl}
\label{leiden}
\and Space Telescope Science Institute, 3700 San Martin Drive, Baltimore, MD 21218, USA
\label{stsci}
\and Universit\'{e} C\^{o}te d'Azur, Observatoire de la C\^{o}te d'Azur, CNRS, Laboratoire Lagrange, F-06304 Nice, France
\label{oca}
\and Universit\'{e} Grenoble Alpes, CNRS, IPAG, 38000 Grenoble, France
\label{grenoble}
\and Department of Physics and Astronomy, University of Victoria, Victoria, BC, V8P 5C2, Canada
\label{victoria}
\and NRC Herzberg Astronomy and Astrophysics, 5071 West Saanich Rd, Victoria, BC, V9E 2E7, Canada
\label{herzberg}
\and Space Science Institute, 4765 Walnut St, Boulder, CO 80301, USA
\label{boulder}
\and Instituto de Astronom\'{i}a, Universidad Nacional Aut\'{o}noma de M\'{e}xico, Apdo. Postal 70264, Ciudad de M\'{e}xico, 04510, México
\label{unam}
\and Max-Planck-Institut f\"{u}r Astronomie, K\"{o}nigstuhl 17, D-69117 Heidelberg, Germany
\label{mpia}
\and INAF, Osservatorio Astrofisico di Arcetri, Largo Enrico Fermi 5,
50125 Firenze, Italy
\label{inaf}
\and LESIA, Observatoire de Paris, PSL, CNRS, Sorbonne Universit\'{e}, Universit\'{e} de Paris, 5 place Janssen, 92195 Meudon, France
\label{lesia}
\and European Southern Observatory, Karl-Schwarzschild-Stra{\ss}e 2,
85748 Garching, Germany
\label{eso}
\and Aix-Marseille Universit\'{e}, CNRS, CNES, LAM, Marseille, France
\label{marseille}
}

\date{Received ?; accepted ?}

\abstract
{Transition disks have large central cavities that have been resolved by imaging surveys during recent years. Cavities and other substructures in circumstellar disks are often interpreted as signposts to massive companions. Detecting companions at small angular separations is challenging with coronagraphic imaging observations.}
{We aim to search for stellar and substellar companions in the central regions of transition disks. Such companions could be responsible for the large dust-depleted cavities. We want to determine if these disks might be circumbinary in their nature, similar to the HD\,142527 system.}
{We observed four systems, HD\,100453, HD\,100546, HD\,135344\,B, and PDS\,70, with the sparse aperture masking mode of VLT/SPHERE, also leveraging the star-hopping method with the adaptive optics system. We extracted the complex visibilities and bispectra from the $H2$ and $H3$ imaging data. A binary model was fit to the closure phases to search for companions and estimate detection limits. For validation, we also analyzed four archival datasets of HD\,142527 and inferred the orbital elements and atmospheric parameters of its low-mass stellar companion.}
{We have not detected any significant point sources in the four observed systems. With a contrast sensitivity of $\approx$0.004, we can rule out stellar companions down to $\approx$2~au and partially explore the substellar regime at separations $\gtrsim$3--5~au. The analysis of HD\,142527\,B revealed that its projected orbit is aligned with dust features in the extended inner disk and that the mutual inclination with the outer disk is close to coplanar for one of the two solutions. Atmospheric modeling confirms the low-gravity and slightly reddened spectral appearance ($\Teff \approx 3300$~K, $\log\,g \approx 3.7$, and $A_V \approx 0.7$). The inferred and derived bulk parameters ($\log\,L_\ast/L_\odot \approx -0.65$, $M_\ast \approx 0.4$~$M_\odot$, and $R_\ast \approx 1.46$~$R_\odot$) are in agreement with dynamical constraints and evolutionary tracks.}
{In contrast to HD\,142527, we find no evidence that a close-in stellar companion is responsible for the resolved disk features of HD\,100453, HD\,100546, HD\,135344\,B, and PDS\,70. Instead of a dynamical effect by a stellar companion, the formation of giant planets or even low-mass brown dwarfs could be shaping the innermost environment ($\lesssim$20~au) of these circumstellar disks, as is the case with the planetary system of PDS\,70.}


\maketitle

\section{Introduction}
\label{sec:introduction}
A ubiquity of substructures have been spatially resolved in protoplanetary disks, both at (sub)millimeter and infrared wavelengths (see reviews by \citealt{andrews2020,benisty2023}). Many of these features have been linked to forming planets \citep[e.g.,][]{zhang2018}, although alternative scenarios have been proposed \citep[e.g.,][]{riols2019}. Disks with dust-depleted inner regions were initially identified by a reduced infrared excess in their spectral energy distributions \citep[SEDs;][]{strom1989,skrutskie1990}. The presence of central cavities in transition disks were later confirmed with high-resolution imaging observations \citep[e.g.,][]{andrews2011,vandermarel2016}. Such systems are commonly thought to host multiple planets \citep{dodsonrobinson2009} or stellar companions \citep{calcino2020} that would carve deep gaps. Historically they have been referred to as transition disks, although it is debated if they are actually transitioning a particular evolutionary phase or if they are just massive, bright disks with an above-average lifetime \citep{garufi2018}. In addition to substructures, localized shadows that are cast by a misaligned inner disk have been detected in several systems \citep{marino2015,benisty2017}. Some shadows are variable which hints at dynamical perturbations on short timescales \citep{stolker2016b,stolker2017}. Shadows and spiral arms seen in scattered light correlate with a high near-infrared (NIR) excess, which also points to a tumultuous inner environment \citep{garufi2018}.

The HD\,142527 system has been emblematic for many of such disk features, including a wide gap that separates the inner and outer disk \citep{verhoeff2011}, spiral-like structures in the far outer disk \citep[e.g.,][]{fukagawa2006,canovas2013,avenhaus2014,rodigas2014,christiaens2014}, an azimuthal asymmetry in the emission from large dust grains \citep{fukagawa2013,casassus2015}, a strongly misaligned inner disk \citep{marino2015}, and variable scattered light from the inner dust structures \citep{avenhaus2017}. Interestingly, a low-mass stellar companion was discovered by \citet{biller2012} within the central cavity of this disk. This M-dwarf companion was detected between 2012 and 2018 at separations in the range of $\approx$44--90~mas (see Table\,4 by \citealt{balmer2022}). The disk of this system is therefore a circumbinary disk. The M-dwarf companion may not have fully formed from the disk, but the object is accreting so it started out at a lower mass than currently observed \citep{close2014,cugno2019}. All of the aforementioned disk characteristics can be explained by the orbital configuration and mass of the inner companion, as was shown by \citet{price2018} in a comprehensive modeling study. 

While the low-mass companion of HD\,142527 was discovered a decade ago, the detection of gap-carving companions in other transitions disks has remained cumbersome. The exception is the PDS\,70 system in which two giant planets have been discovered at small angular separations ($\approx$170--220~mas) from their star \citep{keppler2018,haffert2019}, which enabled a detailed orbital and atmospheric characterization \citep[e.g.,][]{mueller2018,mesa2019,stolker2020b,wang2021}. Since many transition disks have resolved features that have similarities with HD\,142527, this brings up the question of whether other transition disk systems are circumbinary disks or planet-hosting disks. The density, temperature, and velocity structure of a disk in a binary system can be rather different from that of a single-star system. Therefore, to fully understand the structure and evolution of these disks, it is important to search for any low-mass companions. However, probing the central cavities with conventional imaging techniques is challenging due to limited sensitivity at the smallest separations, leading to weak constraints on the presence of companions \citep[e.g.,][]{asensio2021,rich2022}. 

Aperture masking interferometry (AMI) enables high-resolution measurements with an intermediate-contrast at angular separations around the diffraction limit \citep[e.g.,][]{ireland2016}. Therefore, this technique probes a part of the companion mass versus separation space that cannot be reached by regular adaptive optics imaging observations due to use of coronagraphs, self-subtraction effects, and residual speckle noise. AMI has proven to be a successful technique for the detection and characterization of low-mass companions and circumstellar environments at separations of several resolution elements from the star \citep[e.g.,][]{hinkley2015,willson2016,claudi2019}. In particular it is suitable for bright targets since the aperture mask strongly reduces the throughput. Several authors have used AMI to study stellar multiplicity and the brown dwarf desert in moving groups and star-forming regions \citep[e.g.,][]{kraus2008,kraus2011,evans2012,cheetham2015b}.

Early observations with AMI typically reached contrasts of a few hundredths at NIR wavelengths. At higher Strehl ratios, AMI observations are able to probe the planetary-mass regime with limiting contrast of $\sim$10$^{-3}$ \citep[e.g.,][]{hinkley2011,cheetham2019,willson2019}, therefore entering parameter space that cannot be reached by other techniques. A dedicated search for close-in companions in transition disks was carried out by \citet{ruiz2016}, who detected seven binary candidates with AMI among a sample of 24 targets. From their statistical analysis of transition disk targets, which also included 11 additional targets of which the binarity had already been confirmed or ruled out, the authors calculated that a fraction of $0.38 \pm 0.09$ objects with SEDs previously identified as transition disks are in fact circumbinary disks. This suggests that a significant fraction of young circumstellar environments might be shaped by a tidal interaction with an inner stellar companion.

In this work, we investigate the innermost regions ($\approx$2--20~au) of four transition disks, around HD\,100453, HD\,100546, HD\,135344\,B, and PDS\,70, to search for low-mass stars and brown dwarf companions. We aim to answer the question if these systems are circumbinary disks. This also helps to determine whether some of the resolved disk features can be linked to the presence of a (sub)stellar companion instead of other disk evolution processes. To achieve this goal, we used the sparse aperture masking mode of the SPHERE instrument at the Very Large Telescope to investigate separations around the diffraction limit. We also processed several archival datasets of HD\,142527\,B (including an unpublished epoch from 2019) to test our data reduction and calibration tools, and we analyzed the astrometry and (spectro)photometry to characterize the orbit and atmosphere of this low-mass companion.

All targets are young stars in the Sco-Cen association and host a circumstellar disk that has been classified as a transition disk. Specifically, HD\,100453 and HD\,100546 in Lower Centaurus Crux, and HD\,135344\,B, PDS\,70, and HD\,142527 in Upper Centaurus Lupus. Their ages are relevant for converting detection limits into companion masses. We have adopted the ages of the four Herbig stars from \citet{vioque2018}: $6.5^{+0.5}_{-0.5}$~Myr for HD\,100453, $5.5^{+1.4}_{-0.8}$~Myr for HD\,100546, $8.9^{+0.4}_{-0.9}$~Myr for HD\,135344B, and $6.6^{+0.3}_{-1.5}$~Myr for HD\,142527. For the T\,Tauri star in the sample, PDS\,70, we adopted the age of $5.4 \pm 1.0$~Myr from \citet{mueller2018}. Our main objective is to search for low-mass companions in the central cavities of those systems. The cavity radii, $R_\mathrm{Cav}$, have been systematically inferred from millimeter continuum emission by \citet{francis2020}: 0\ffarcs29 ($=$30~au) for HD\,100453, 0\ffarcs23 ($=$25~au) for HD\,100546, 0\ffarcs38 ($=$52~au) for HD\,135344\,B, 0\ffarcs65 ($=$74~au) for PDS\,70, and 1\ffarcs18 ($=$185~au) for HD\,142527. The millimeter cavities span therefore the full field-of-view of the AMI observations, but micron-sized dust grains are typically present within those cavities since their dynamics are coupled to the gas.

The paper is organized as follows: Sect.~\ref{sec:observations} presents the observations and the procedures for the reduction and calibration of the data. Sect.~\ref{sec:searching_companions} presents the analysis on the search for companions and estimation of detection limits, and Sect.~\ref{sec:characterization_hd142527b} presents the orbital and atmospheric analysis of HD\,142527\,B. Finally, we discuss our findings in Sect.~\ref{sec:discussion} and conclude in Sect.~\ref{sec:conclusions}.

\begin{figure*}
\centering
\includegraphics[width=\linewidth]{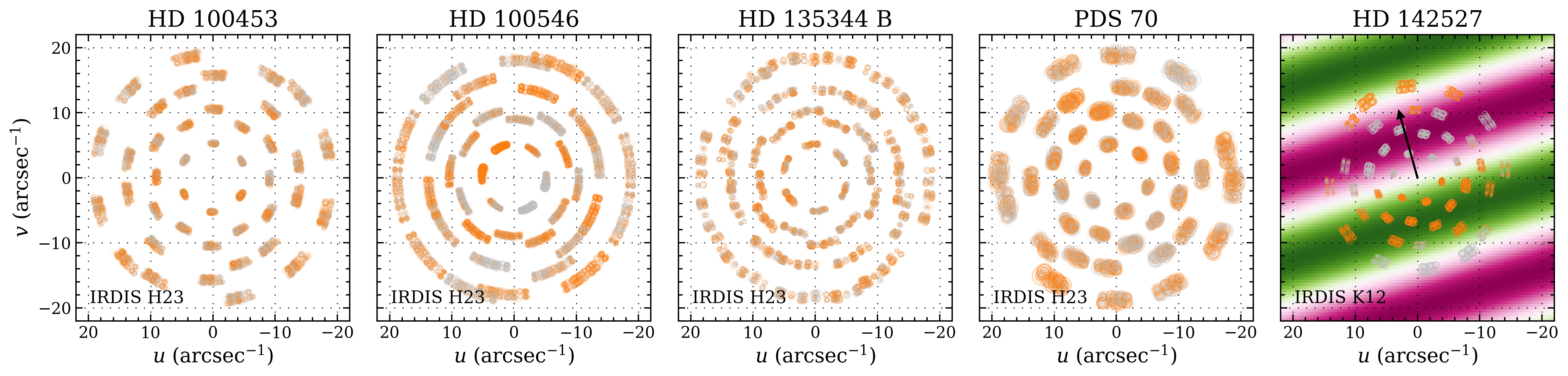}
\caption{Coverage of the $(u,v)$-plane for the observations of HD\,100453, HD\,100546, HD\,135344\,B, PDS\,70, and HD\,142527 (\emph{from left to right}), with $K12$ for HD\,142527 and $H23$ for all other targets. The data points are shown with \emph{circular markers}, with the shortest wavelength of the dual-band filters corresponding to the slightly larger baselines. The \emph{marker size} gives an indication of the phase in the Fourier domain, while the \emph{orange} and \emph{gray} colors corresponds to a positive and negative sign of the phase, respectively. The background colors in the panel of HD\,142527 show the phase pattern from the retrieved model parameters (see Sect.\,\ref{sec:hd142527_astrometry_photometry}) and the arrow points in the direction of HD\,142527\,B.}
\label{fig:uv_coverage}
\end{figure*}

\section{Observations and data reduction}
\label{sec:observations}

\subsection{Sparse aperture masking with VLT/SPHERE}
\label{sec:sphere_sam}

The observations were carried out with the Spectro-Polarimetric High-contrast Exoplanet REsearch (SPHERE; \citealt{beuzit2019}) instrument that is mounted on Unit Telescope 3 of the Very Large Telescope (VLT) at the Paranal Observatory. We used the sparse aperture masking (SAM; \citealt{cheetham2016}) mode in combination with the Infra-Red Dual-beam Imaging and Spectroscopy (IRDIS; \citep{dohlen2008}) camera and the integral field spectrograph (IFS; \citealt{claudi2008}). A nonredundant 7-hole aperture mask (\emph{N\_SAM\_7H}) was placed in the optical path, thereby transforming the single pupil into an interferometer. SPHERE observations employ the extreme adaptive optics (AO) system \citep{fusco2006}, enabling high-contrast measurements at small angular separations. The narrowband $H2$ ($\lambda_\mathrm{c} = 1.59$~$\mu$m) and $H3$ ($\lambda_\mathrm{c} = 1.67$~$\mu$m) filters were used for the dual-band imaging with the IRDIS arm and low-resolution ($R \approx 50$) spectra were obtained across the $YJ$ bands ($\lambda \approx 0.95$--$1.68$~$\mu$m) with the IFS. The pupil of the telescope was stabilized such that speckles remain quasi-static on the detector. At the same time, this conveniently improves the coverage of the $(u,v)$-plane by the parallactic rotation.

The data were obtained on the nights of 2021~Jan~31, Feb~25, Mar~16, Jul~15, and 2022~Jul~01 (program ID: 105.2067.001). Two observing blocks (OBs) of 1~hr each were executed for HD\,100453 and HD\,100546, whereas 1~hr of telescope time was used for HD\,135344\,B and PDS\,70. The detector integration time (DIT) was 0.84~s for the three Herbig stars and 2~s for the T\,Tauri star PDS\,70. Each exposure consisted of about 30 integrations and every four exposures alternated between a calibrator and the science target, reducing the effective on-source integration time to 10--20\,min per OB. For changing between science and calibration target, we benefited from the \emph{star-hopping} method of SPHERE, such that the AO loop did not need to be reoptimized. This ensured a similar AO correction for both targets, and therefore a higher similarity of their point spread functions (PSFs), as well as an improved time efficiency of the observations \citep{wahhaj2021}. The overhead when changing between science a calibration target was typically a few minutes. For the calibrators, we used the \texttt{SearchCal} tool \citep{Chelli2016} to select single stars in the proximity of the science targets and with a similar NIR brightness and color. All observations had been executed in quite stable conditions with a typical seeing of $\approx$1\ffarcs0 and a turbulence coherence time of $\approx$5--8\,ms.

We also analyzed four archival datasets of HD\,142527, to test the data reduction and calibration, and to confirm or update the orbital and atmospheric constraints of the low-mass companion in this system. Three of the datasets (programs ID: 60.A-9800(S), 198.C-0209(G), 1100.C-0481(F)) have been published by \citet{claudi2019} so we refer the reader to their work for details on the observations. The fourth dataset (program ID: 1100.C-0481(N)) has not yet been published. It was obtained on the night of 2019\,May\,17 with a DIT of 0.84~s and total integration time of 8\,min. An equal number of exposures were obtained for the calibration targets that were observed both before and after the sequence on HD\,142527. Observing conditions were stable with a seeing of $\approx$0\ffarcs8--0\ffarcs9 and coherence time of $\approx$2--3\,ms.

\subsection{Image processing}
\label{sec:image_processing}

The PSF of the aperture masking observations is an interferogram which is imaged by the IRDIS and IFS cameras. The fringes are generated by the subapertures in the pupil mask, and contain the amplitude and phase information of the underlying source brightness distribution. The specific 7-hole mask that we used formed a set of 21 nonredundant baselines. With aperture masking inteferometry, we can probe down to separations of $0.5 \lambda/B_\mathrm{max}$ with $\lambda$ the wavelength and $B_\mathrm{max}$ the maximum baseline between two holes in the nonredundant mask (i.e., super-resolution since $B_\mathrm{max} \approx D$, with $D$ the primary mirror diameter).

The processing of the images was done with the \texttt{vlt-sphere}\footnote{\url{https://github.com/avigan/SPHERE}} package that provides a Python wrapper to the standard \emph{EsoRex} recipes for SPHERE \citep{vigan2020}. For the IRDIS data, the basic preprocessing involved subtraction of the sky background, flat field normalization, bad pixel correction, a coarse centering with pixel precision (so without interpolation), and cropping of the images. The pipeline also calculates the parallactic angles, which are required for determining the correct orientation of the field of view since the observations were executed with pupil tracking.

\subsection{Extraction and calibration of complex observables}
\label{sec:extraction_calibration}

The interferometric observables were extracted and calibrated with the \texttt{AMICAL}\footnote{\url{https://github.com/SAIL-Labs/AMICAL}} pipeline \citep{soulain2020}. We started the procedure by removing $3\sigma$ outliers with a low central flux (e.g., due to reduced AO correction) and cropping the images to 251 by 251 pixels. Next, images were apodized to limit the contribution of high-frequency read-out noise and reduce numerical edge artifacts when calculating the Fourier transform later on. Specifically, the images were windowed by scaling with a super-Gaussian function of the following form that is implemented in \texttt{AMICAL}:
\begin{equation}
\label{eq:super_gaussian}
A(r) = a\,\exp{\left[-2^{2m-1}\,\log{(2)}\,{\left(\frac{r}{w}\right)^{2m}}\right]}^2
\end{equation}
where $r$ is the radius from the image center, $w$ is the full width at half maximum (FWHM), which we set to 172~pixels for all datasets. The amplitude is set internally to $a=1$ and the exponent to $m=3$. This apodization is characterized by a flat top and a sharp tapering at the separation of the FWHM. The images were then Fourier transformed such that the complex visibilities could be extracted for the 21 baselines. The splodge positions are determined from the aperture positions in the mask and pixel scale of the IRDIS camera. Each splodge covers multiple pixels in the Fourier plane so the pipeline extracts the visibilities and bispectra by calculating a weighted average over a subsample of pixels (see \citealt{soulain2020} for details).

The squared visibilities and closure phases were then computed while averaging over subsets of two images. For a 7-hole mask, there are 21 visibilities and 35 closure phases. These quantities were subsequently calibrated by using the dedicated calibration data that has been processed in an identical manner as the science data. For the calibration, we selected the group of exposures that were obtained closest in time to the science data and divided (subtracted) the visibility amplitudes (closure phases) with a noise-weighted combination of the observables extracted from the calibration data to remove (part of) the systematics from the science data. The extraction and calibration was done for each of the two filters separately. The coverage of the $(u,v)$-plane is shown in Fig.\,\ref{fig:uv_coverage} that combines the two filters used for each target. The arcs trace the rotation of the aperture mask with respect to the sky. There were two datasets obtained for HD\,100453 and HD\,100546 with a small overlap in parallactic angles that can be noticed in the figure.

\subsection{SpeX observations of HD\,142527}
\label{sec:spex_observations}

Flux measurements of HD\,142527\,B are relative to the brightness of the (unresolved) star-disk system. Since HD\,142527 is a young star with a circumstellar disk, it is important to account for the NIR excess so we use an empirical spectrum of HD\,142527 for the absolute flux calibration of the companion. The spectrum was obtained with the SpeX spectrograph \citep{rayner2003} on NASA's Infrared Telescope Facility (IRTF). The system had been observed on UT\,2013\,May\,15 with the SXD grating ($0.7$--$2.4$~$\mu$m) and a slit width of 3\ffarcs0 ($R \approx 75$). The \texttt{Spextool} \citep{cushing2004} was used for the data reduction and spectral extraction. The package also applies a telluric correction and flux calibration \citep{vacca2003}, for which calibration spectra had been obtained of the A0V-type star HD\,130163.

\begin{figure}
\centering
\includegraphics[width=\linewidth]{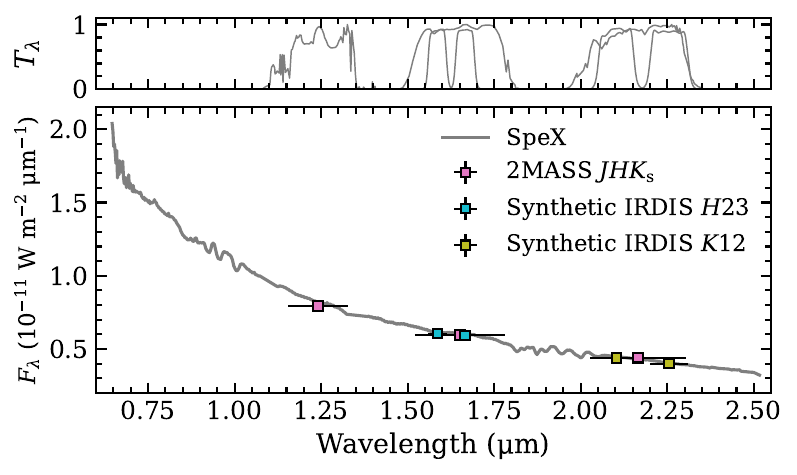}
\caption{SpeX spectrum of HD\,142527 at $R \sim 75$. The \emph{colored markers} are the 2MASS $JHK_\mathrm{s}$ fluxes and synthetic IRDIS $H23$ and $K12$ photometry. The \emph{horizontal error bars} are the full width at half maximum of the filter transmission profiles, which are shown in the \emph{top panel}.}
\label{fig:spex}
\end{figure}

The relative fluxes of the cross-dispersed spectra had been accurately calibrated but the absolute flux was impacted by variable conditions between observations of the science and calibrator target. We therefore fit and scaled the SpeX spectrum to the 2MASS $JHK_\mathrm{s}$ fluxes. This provided a good fit with residuals at the $\sim$1--2$\sigma$ level of the 2MASS uncertainties. The SpeX spectrum of HD\,142527, including the applied best-fit scaling ($\approx$1.25), is shown in Fig.\,\ref{fig:spex}. The consistency between the spectral slope of the SpeX spectrum and the 2MASS $JHK_\mathrm{s}$ colors suggests that the NIR is likely not impacted by any variability, since such an effect is expected to show up as a differential slope between SpeX and 2MASS.

From the calibration spectrum, we then computed synthetic IRDIS and IFS fluxes. To do so, we used the \texttt{species}\footnote{\url{https://github.com/tomasstolker/species}} toolkit \citep{stolker2020a}, with which we determined the following magnitudes for the IRDIS dual-band filters: $5.81 \pm 0.03$ in $H2$, $5.65 \pm 0.03$ in $H3$, $5.07 \pm 0.02$ in $K1$, and $4.89 \pm 0.02$ in $K2$, using a flux-calibrated spectrum of Vega \citep{bohlin2014} and setting its magnitude to 0.03 for all filters. The synthetic fluxes are also shown in Fig.\,\ref{fig:spex}, with the uncertainties being dominated by the 2MASS photometry. We also computed a synthetic IFS spectrum of HD\,142527 by smoothing the SpeX spectrum to $R = 30$ and resampling to the wavelength solution of the instrument.

\section{Searching for low-mass companions}
\label{sec:searching_companions}

\subsection{Detection maps from fitting a binary model}
\label{sec:chi2_map}

The main goal of this work is to search for stellar, substellar, or even planetary-mass companions at the smallest separations possible in the central cavities of transition disks. To find such objects, we fit the extracted closure phases with a binary model. The complex visibilities of the model are calculated as
\begin{equation}
\label{eq:binary_vis}
V(u,v) = \frac{1 + f\exp{\left[-2\pi i(u\alpha+v\beta)\right]}}{1+f},
\end{equation}
where $f$ is the companion-to-star flux ratio, $u$ and $v$ are the coordinates in the Fourier plane, and $\alpha$ and $\beta$ are the angular coordinates of the companion relative to the star. These were then converted to squared visibility amplitudes, $|V|^2$, and closure phases for the baselines of the observations, such that data and model can be compared. Although the data were obtained with narrowband filters, some smearing is to be expected in the complex observables. To account for that, we calculated the model visibilities for three wavelengths that are equally spaced across the filter bandpasses, and continue with the wavelength-averaged visibilities.



To identify any companions in the closure phases, we calculated $\chi^2$ detection maps using \texttt{fouriever}\footnote{\url{https://github.com/kammerje/fouriever}}, a toolkit for analyzing various types of interferometric data \citep{kammerer2023}. The field-of-view of the observations is $\lambda_\mathrm{max}/B_\mathrm{min}$, with $\lambda_\mathrm{max}$ the longest wavelength of the two dual-band filters and $B_\mathrm{min}$ the shortest baseline of the aperture mask. The smallest separation that can be resolved with interferometry is given by the Michelson criterion, $\delta\theta = 0.5 \lambda/D$, with $\lambda$ the wavelength and $D$ telescope aperture. In the $H$ band, this corresponds to a separation of $\approx$20~mas, so a factor $\approx$2.4 higher resolution than with regular imaging. An accurate calibration could in principle permit the detection of features at even smaller separations, so we loosely test for companions down to separations as small as 10~mas.

The detection maps were calculated from the inner separation of 10~mas up to 190~mas and 250~mas for the $H23$ and $K12$ data, respectively. Specifically, the maps are constructed by calculating what the contrast and detection significance would be if a companion were to be present at a certain location. The best-fit solutions within the separation range are determined by initializing a grid of least-squares minimizations with a resolution of 10~mas and letting the companion contrast and position converge toward the local minimum. This ensures that the global best-fit solution is found albeit the $\chi^2$ map typically showing substructures of multiple local minima in the presence of a companion. These aliases of the true detection are present due to the incomplete $(u,v)$-plane coverage of the observations. Since the closure phases are symmetric, we set all nonphysical solutions with negative companion flux to zero in the detection maps. Before calculating the maps, we used \texttt{fouriever} to estimate the covariances due to overlapping baselines between closure phase triangles \citep{kammerer2019,kammerer2020}, based on the uncertainties estimated with \texttt{AMICAL}.

\begin{figure*}
\centering
\includegraphics[width=0.86\linewidth]{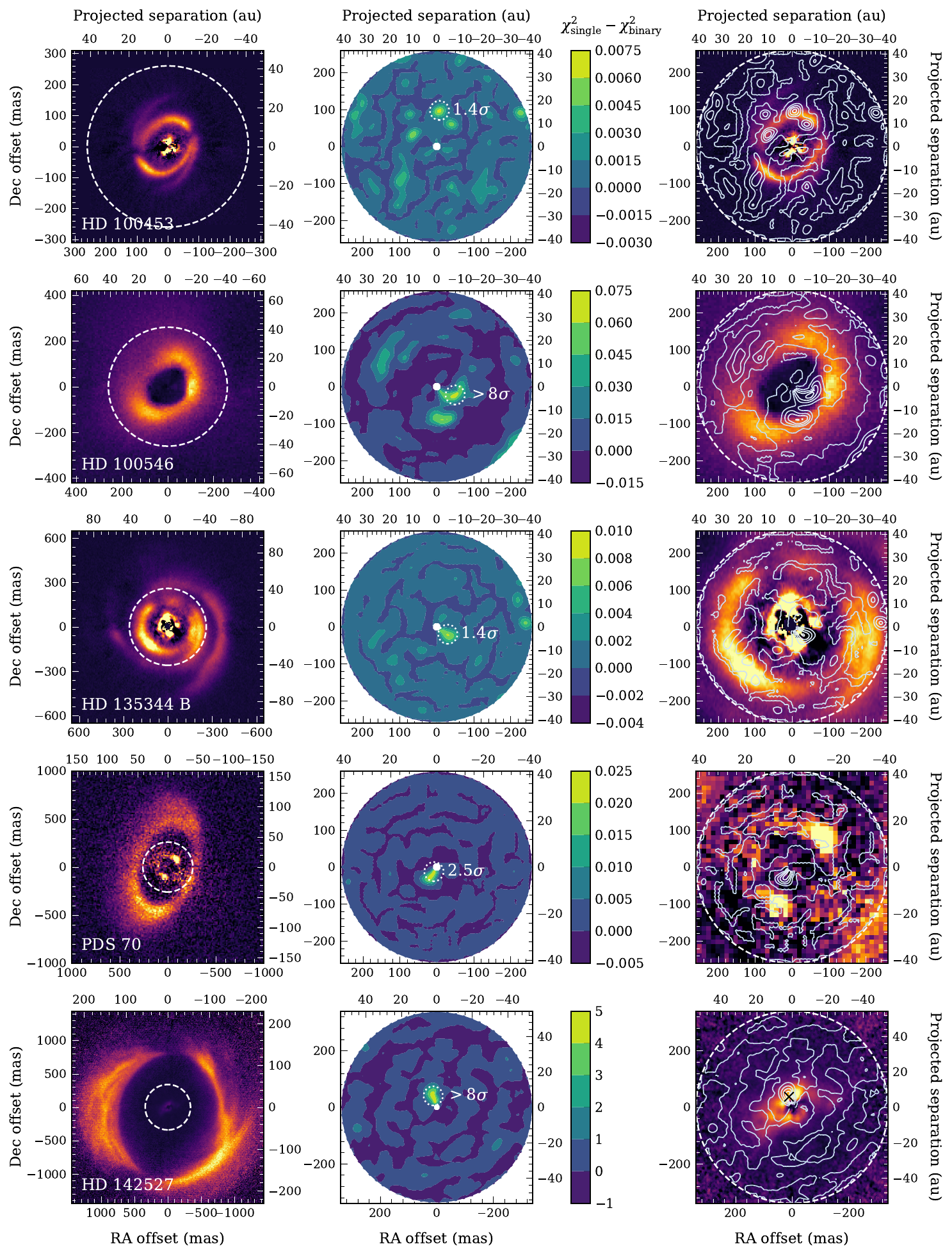}
\caption{Detection maps for companions. The \emph{left column} shows full images of the disks, the \emph{central column} the $\chi^2$ maps, and the \emph{right column} a zoom to the field of view of the AMI observations with \emph{contours} showing the $\chi^2$ levels. The \emph{dashed circles} correspond to the field of view. The $\chi^2$ maps are calculated as the difference in the goodness-of-fit statistic between a model without companion and the minimum $\chi^2$ of a model with a companion. The position with the highest confidence level is encircled in the maps of the central column. It is important to note that the dynamical range of the colorbar is different between targets. See main text in Sect.\,\ref{sec:chi2_map} for details on the interpretation of the contours. The \emph{black cross} in the bottom right panel is placed at the best-fit position of HD\,142527\,B. The disk images are in polarized scattered light, all obtained in the optical with ZIMPOL, except PDS\,70 in the near-infrared with IRDIS. From \emph{top to bottom} the images have been adopted from: \citealt{benisty2017}, \citealt{garufi2016}, \citealt{stolker2016b}, \citealt{vanholstein2021}, \citealt{avenhaus2017}.}
\label{fig:chi2maps}
\end{figure*}

The detection maps are presented in Fig.~\ref{fig:chi2maps}. The maps show the difference in the reduced chi-squared statistic, $\chi^2_\nu = \chi^2/\nu$ with $\nu$ the number of degrees of freedom, between a single-star model and the best-fit binary model. The $\chi^2$ maps were converted into confidence maps by adapting the procedure that is outlined in \citet{absil2011} (see also \citealt{gallenne2015}). To summarize, we calculated $\chi^2_\nu (f=0)$ at each position, $(x, y)$, in the detection map. This is the $\chi^2$ statistic of the single-star model for which the companion-to-star flux ratio is set to zero. That $\chi^2_\nu$ value is normalized by the value from the best-fit binary model at that same position, $\mathrm{min}(\chi^2_\nu(f))$, such that $\chi^2_\nu = 1$ when the best-fit binary model yields a similar statistic as the single-star model. The normalization is required because the measurements are limited by systematic uncertainties instead of photon noise. We then computed the probability, $P_0$, for the measured statistic to be equal to or larger than $\chi^2_\nu(f=0)$:
\begin{equation}
\label{eq:chi2_prob}
P_0 = 1 - \mathrm{CDF}_\nu\left(\frac{\nu\chi^2_\nu(f=0)}{\mathrm{min}(\chi^2_\nu(f)} \right)
\end{equation}
where $\mathrm{CDF}_\nu$ is the $\chi^2$ cumulative distribution function with $\nu$ degrees of freedom. The hypothesis of the single-star can be rejected if the probability, $P_0$, is below a predefined threshold for the detection significance. We can now calculate the confidence level of a binary detection in number of $\sigma$:
\begin{equation}
\label{eq:chi2_sigma}
\mathrm{CL} = \sqrt{\mathrm{PPF}_{\nu=1}\left(1-P_0\right)}
\end{equation}
where $\mathrm{PPF}_{\nu=1}$ is the $\chi^2$ percent point function with one degree of freedom.

For each target, the feature with the highest confidence level is indicated in the central column of Fig.~\ref{fig:chi2maps}. Maps were calculated separately for each observation and filter, as well as combining all IRDIS data of a target. The latter is shown in the figure. We inspected the maps by eye to determine if any potential astrophysical sources are detected within the targets' cavities. For reference, we first studied the detection map of HD\,142527. There is one prominent, high signal-to-noise ratio (S/N) feature see in the map (see bottom row in Fig.\,\ref{fig:chi2maps}). The position and contrast with the minimum $\chi^2$ is selected as the parameters of HD\,142527\,B, which will be more carefully extracted in Sect.\,\ref{sec:hd142527_astrometry_photometry}. There is also a pattern of several weaker features (although also with high significance) in the detection map of HD\,142527. This is caused by the effect of the incomplete $(u,v)$ coverage that was mentioned earlier.

Now that we know what signal to look out for from a companion, we inspect the detection maps of the four observed targets. There were no obvious sources comparable to HD\,142527\,B within the central cavities that are seen in scattered light. Some features are present, but typically at low confidence ($\approx$1--2$\sigma$) and not consistently seen when comparing $\chi^2$ maps from different filters and/or different observing nights. Therefore, we expect such features to be spurious signals caused by remaining systematics in the data. The visibility amplitudes were affected by even stronger systematic uncertainties than the closure phases. Including $|V|^2$ in the fit reduced the significance of a detection, which we tested with the data of HD\,142527. We therefore only analyzed the closure phases and excluded the amplitudes from all the analyses. This implies that the analysis is blind to any emission that is point-symmetric with respect to the host star.

Although there were not any obvious companions detected, we provide a few details on the detection maps of the individual targets:

\begin{itemize}
  \item \emph{HD\,100453}: Low S/N features are present throughout the field-of-view with a few coinciding with the gap edge of the protoplanetary disk. We note however that the image shows polarized intensity while in total intensity the scattered light flux is highest on the near side of the disk that is located in SW direction. With our observations we would be sensitive to scattered light features in total intensity so an effect of the gap edge would more likely show up in SW direction.

  \item \emph{HD\,100546}: There are two features with a high S/N that coincide with the gap edge on the near side of the disk, which is in SW direction. There is also a more elongated feature in the map at the opposite side of the star, also coinciding with the gap edge. We expect that these features are caused by scattered light and not associated with any companion. The near side of the disk will be brightest due to forward scattering, which is indeed where the high S/N features are seen. One of these features appears to be located slightly inward of the gap edge though. The reason for this is unclear but we expect that it could be an artificial effect that is caused by the correlation between the contrast and angular separation at locations close to and below the diffraction limit. At such small separations, the sinusoidal phase modulation from a binary source is not fully sampled, while that would be required to separately constrain the contrast and separation \citep[see e.g.,][]{martinache2010,willson2016}.

  \item \emph{HD\,135344\,B}: There are not any significant sources detected. The highest confidence level is only $1.4\sigma$ for a feature that is located in SW direction at a separation of $\approx$30 mas and contrast of $\approx$7~mag. It is likely a spurious signal or could be scattered light from an extended inner component. We mention it here because it still stands out in the detection map. The gap edge shows a bit of an azimuthal brightness variation in polarized light but the disk is seen a small inclination. Therefore, signal from the gap edge may not show up in the closure phases.

  \item \emph{PDS\,70}: Similar to HD\,135344\,B, there is one close-in feature that stands out in the detection map but again with a low confidence level. It is aligned with the major axis of the inclined inner disk, of which the outer parts are visible in polarized scattered light, just beyond the coronagraph mask (see Fig.~\ref{fig:chi2maps}). We therefore expect that this feature traces scattered light from the inner disk and not emission from a companion. The detection of the inner disk in the AMI data is at the smallest separation at which we are sensitive, which corresponds to the brightest part of the irradiated disk.

  \item \emph{HD\,142527}: As already explained earlier, the feature in the detection map with the highest confidence ($\gg$8$\sigma$) corresponds to the position of the known low-mass companion, HD\,142527\,B (see black cross in the bottom right panel of Fig.~\ref{fig:chi2maps}).
\end{itemize}

\subsection{Sensitivity and upper limits on companion masses}
\label{sec:sensitivity_limits}

The detection limits were computed with \texttt{fouriever} from the calibrated data products for the IRDIS $H2$ and $H3$ filters (or $K1$ and $K2$ in case of HD\,142527) separately, as well for the two filters combined. Given the small wavelength difference between the two filters, we assume an equal brightness of potential companions at those wavelengths. This may no longer be a valid assumption for T-type objects, but we do not expect to reach such late type objects at the ages of the observed systems.

The limits were estimated from the closure phases by analytically injecting companion sources and optimizing the detection significance to a confidence level of $5\sigma$, again using Eqs.~\ref{eq:chi2_prob} and~\ref{eq:chi2_sigma}. As a reminder, the $5\sigma$ detection limit is defined as the relative companion flux at which the binary model deviates by more than $5\sigma$ from the single-star model, assuming $\chi^2$ statistics. Since this method considers the ratio of the reduced $\chi^2$ statistic of the two cases, it is in principle independent of an under- or overestimation of the uncertainties. Systematic uncertainties could however bias the inference from the data when fitting the binary model, thereby potentially impacting the derived detection limits.

\begin{figure*}
\centering
\includegraphics[width=0.69\linewidth]{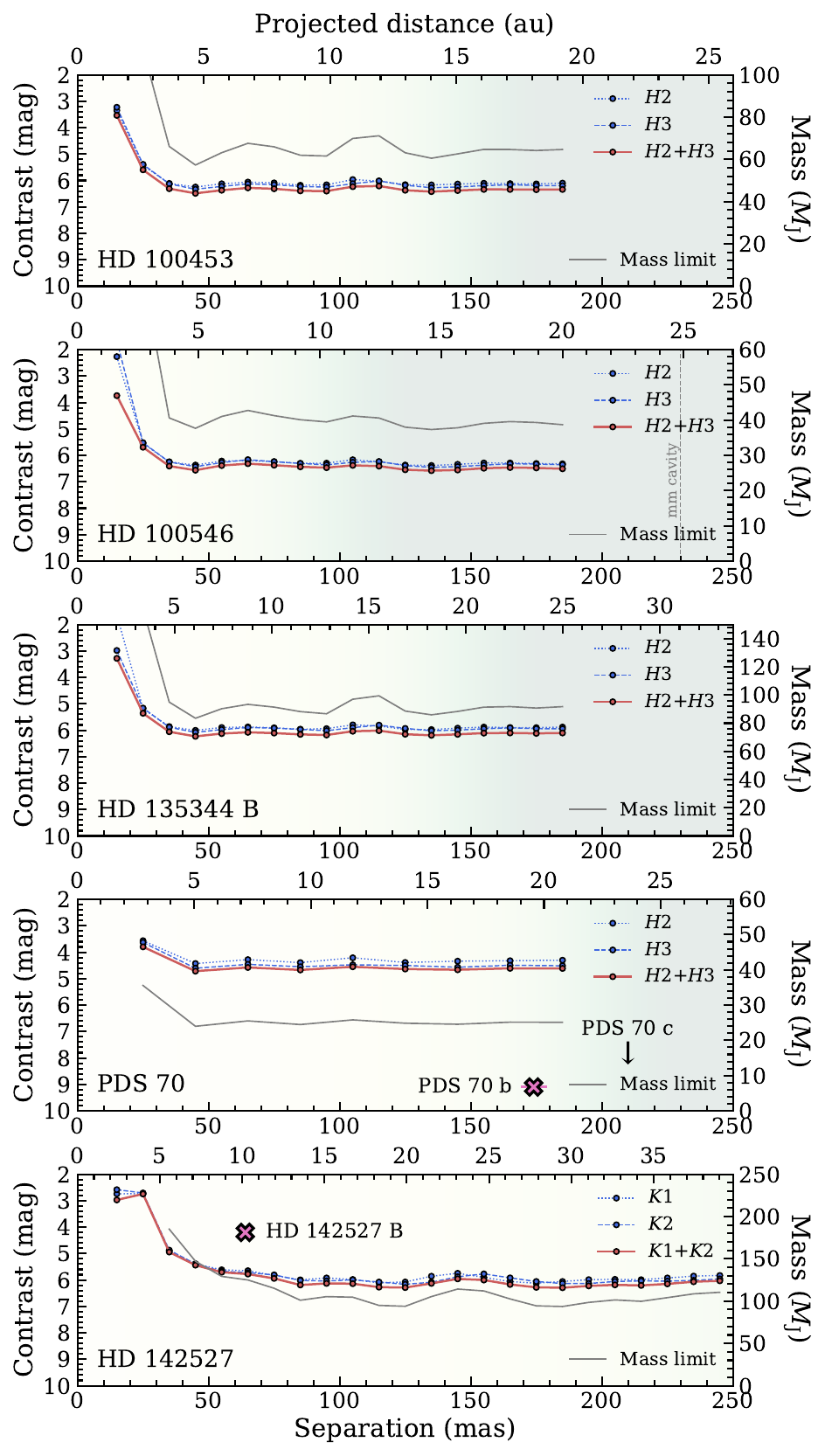}
\caption{Detection limits estimated from the closure phases of the IRDIS dual-band data. The confidence level of the contrast limits is $5\sigma$ and the used narrowband filters are indicated in the legend. From \emph{top} to \emph{bottom}, the panels show the limits for HD\,100453, HD\,100546, HD\,135344\,B, and PDS\,70. The \emph{gray lines} show the corresponding masses at the distance and stellar magnitude of each system, on the scale of the \emph{right axis} (see main text for details). The \emph{green shading} indicates the approximate gap regions that are seen in scattered light images with \emph{dark colors} corresponding to higher fluxes (see also Fig.~\ref{fig:chi2maps}).} The \emph{vertically dashed line} indicates the millimeter cavity radius of HD\,100453 that is within the field-of-view of the observations.
\label{fig:limits}
\end{figure*}

We have calculated detection limits up to the outer edge of the field-of-view, which is $\approx$190~mas and $\approx$250~mas in the $H23$ and $K12$ bands, respectively. As for the detection maps (see Sect.\,\ref{sec:chi2_map}), we explore separations as small as 10~mas (i.e., $\approx$$0.25\lambda/D$) and account for the covariances as mentioned earlier. The azimuthally averaged contrast limits at a confidence level of $5\sigma$ are presented in Fig.\,\ref{fig:limits}. The limits are comparable between the targets, reaching as deep as 6.0--6.5~mag at separations $\gtrsim$40~mas (i.e., $\approx$$\lambda/D$) and show a sharp decrease to a contrast of $\approx$3.5~mag at 10~mas. The contrast limits for PDS\,70 are smaller and reach only $\sim$4.5~mag since this is the faintest star in the sample with $H = 8.8$~mag (compared to $H \approx 6.0$--$6.5$ for the other systems). The panels also show the corresponding mass limits that were interpolated from the isochrones by \citet{baraffe2015}. For the conversion from contrast to mass, we adopted the Gaia DR3 parallaxes \citep{gaia2016,gaia2023} and the 2MASS $H$-band magnitudes \citep{cutri2003}, so ignoring minor color effects between the 2MASS and SPHERE filters.

\section{Orbital and spectral analysis of HD\,142527\,B}
\label{sec:characterization_hd142527b}

\subsection{Astrometry and spectrophotometry with SPHERE/SAM}
\label{sec:hd142527_astrometry_photometry}

\begin{table*}
\caption{Astrometry and photometry of HD 142527 B with SPHERE/SAM.}
\label{table:astrometry_photometry}
\centering
\bgroup
\def\arraystretch{1.25}
\begin{tabular}{L{2cm} C{1.5cm} C{2cm} C{2cm} C{2cm} C{2.3cm} C{3cm}}
\hline\hline
UT date & Filter & Separation & Position angle & Contrast & App. magnitude & Flux\\
 & & (mas) & (deg) & (mag) & (mag) & (W\,m$^{-2}$~$\mu$m$^{-1}$)\\
\hline
\multirow{2}{*}{2015 Jul 03} & IRDIS $H2$ & \multirow{2}{*}{$64.1 \pm 0.7$} & \multirow{2}{*}{$106.7 \pm 0.6$} & $4.23 \pm 0.11$ & $10.03 \pm 0.12$ & $1.28 \pm 0.14$ $\times$ $10^{-13}$ \\
 & IRDIS $H3$ &  &  & $4.13 \pm 0.11$ & $9.78 \pm 0.11$ & $1.35 \pm 0.14$ $\times$ $10^{-13}$ \\
\multirow{2}{*}{2017 May 17} & IRDIS $K1$ & \multirow{2}{*}{$48.7 \pm 0.8$} & \multirow{2}{*}{$75.6 \pm 1.1$} & $4.68 \pm 0.07$ & $9.75 \pm 0.07$ & $6.12 \pm 0.40$ $\times$ $10^{-14}$ \\
 & IRDIS $K2$ &  &  & $4.68 \pm 0.08$ & $9.57 \pm 0.08$ & $5.60 \pm 0.40$ $\times$ $10^{-14}$ \\
\multirow{2}{*}{2018 Apr 14} & IRDIS $K1$ & \multirow{2}{*}{$42.8 \pm 1.8$} & \multirow{2}{*}{$51.7 \pm 2.6$} & $4.61 \pm 0.18$ & $9.68 \pm 0.19$ & $6.50 \pm 1.11$ $\times$ $10^{-14}$ \\
 & IRDIS $K2$ &  &  & $4.55 \pm 0.18$ & $9.44 \pm 0.18$ & $6.26 \pm 1.07$ $\times$ $10^{-14}$ \\
\multirow{2}{*}{2019 May 18} & IRDIS $K1$ & \multirow{2}{*}{$38.0 \pm 0.7$} & \multirow{2}{*}{$16.5 \pm 1.8$} & $4.63 \pm 0.16$ & $9.70 \pm 0.16$ & $6.39 \pm 0.95$ $\times$ $10^{-14}$ \\
 & IRDIS $K2$ &  &  & $4.61 \pm 0.15$ & $9.50 \pm 0.15$ & $5.96 \pm 0.84$ $\times$ $10^{-14}$ \\
\hline
\end{tabular}
\egroup
\end{table*}

In the previous section we determined the point-source sensitivity and derived upper limits on companion masses. We did not detect any new companion. In this section we analyze the astrometry and spectrophotometry of HD\,142527\,B to provide an updated view on its orbit and atmospheric characteristics. We analyzed four datasets that had been obtained between 2015 and 2019. The data from 2015, 2017, and 2018 have already been published by \cite{claudi2019}. As mentioned in Sect.\,\ref{sec:sphere_sam}, we reanalyzed these data to validate the data reduction and calibration procedure. As we will see shortly, the relative astrometry and photometry that we derived is consistent with the results by those authors. The fourth dataset from 2019 has not yet been published and will be analyzed in a similar fashion.

The position and brightness of HD\,142527\,B is measured relative to its host star. From the $\chi^2$ map (see Sect.~\ref{sec:chi2_map}), we took the position and contrast with the highest significance as the approximate values for an exploration of the multidimensional posterior distribution. The statistical inference was done with a binary model (see Eq.\,\ref{eq:binary_vis}) and the nested sampling algorithm of \texttt{MultiNest} \citep{feroz2008}, through the Python interface of \texttt{PyMultiNest} \citep{buchner2014}. The prior space was explored with 200\,live points, while accounting for the covariances in the (Gaussian) likelihood function. With the calculation of the model visibilities, we accounted again for smearing as described in Sect.\,\ref{sec:chi2_map}. We also renormalized the likelihood function because the reduced $\chi^2$ of the detections of HD\,142527\,B indicated that the variances had been underestimated by a factor of $\approx$1.5--2.5 for the four datasets of HD\,142527.

For each observation, the astrometric and photometric measurements were simultaneously carried out for both of the dual-band filters, so fitting one position and two flux contrasts. The posterior distributions are not included but we point out that the 1D marginalized distributions appeared symmetric and Gaussian-like. We adopted the mean and standard deviation of the posterior samples as best-fit values and statistical uncertainties, respectively. The 2D projections derived from the $K12$-band data showed a slight negative correlation between separation and contrast. As pointed out previously, this degeneracy is to be expected at separations close to the diffraction limit. This effect can be noticed in Fig.~\ref{fig:posterior_extraction} of Appendix~\ref{sec:appendix_posteriors}, which shows as an example the posterior distributions of the retrieved binary parameters for the IRDIS dataset of HD\,142527 that was obtained in 2019. Interestingly, while the correlation is seen in all three $K12$ datasets, it is absent in the posterior distribution of the $H23$ data. The reason might the higher resolution in the $H$ band and somewhat larger separation of the companion at that epoch (see Table~\ref{table:astrometry_photometry}). Also, the bandwidth of the $H23$ filters is a factor $\approx$2 smaller compared to $K12$, therefore it is less prone to the smearing effect.

The statistical uncertainties seemed underestimated, even though the (co)variances had been inflated with the fit. We therefore quantified the systematic uncertainty with an injection-retrieval test. Specifically, we first subtracted the actual companion from the closure phases, using a binary model with the best-fit parameters but a negative flux. Next, we injected at 360 evenly spaced position angles an artificial companion with the same contrast and separation as HD\,142527\,B. We then retrieved the parameters and used the distribution of the difference with the injected values as an estimate of the systematic uncertainty. These were then added in quadrature to the error budget. We also included the uncertainty on the true north, $-1.75 \pm 0.08$~deg \citep{maire2016b}, in the error budget of the position angle.

The astrometry and photometry are listed in Table\,\ref{table:astrometry_photometry}. All uncertainties are $1\sigma$ and include both the statistical and systematic components. The error budgets are dominated by the systematic uncertainties. The apparent magnitudes are calculated from the measured contrast of HD\,142527\,B and the synthetic magnitudes of the central star (see Sect.\,\ref{sec:spex_observations}). The magnitudes of the companion were then converted to physical fluxes with the \texttt{species} toolkit \citep{stolker2020a}, again using a flux-calibrated spectrum of Vega \citep{bohlin2014} and setting its magnitude to 0.03 for all filters.

The low-resolution IFS spectra were also extracted with \texttt{fouriever}. We extracted the contrast for the 39 wavelength channels simultaneously from the calibrated closure phases. We used again the nested sampling algorithm to estimate the posteriors of the 41 free parameters, that is, 39 wavelengths and the RA and Dec relative to the central star. We note that we also tested the extraction on each wavelength channel separately. That approach is computationally less expensive but fits a position for each wavelength. This yielded a contrast spectrum that is consistent within 1--2$\sigma$ with the extraction that fit all wavelengths simultaneously.

The contrast spectrum was converted to spectral fluxes by multiplying with the synthetic calibration spectrum derived from the SpeX spectrum (see Sect.\,\ref{sec:spex_observations}). The extracted spectrophotometry is listed in Table\,\ref{table:ifs_spectrum} of Appendix\,\ref{sec:appendix_spectrum}. For the companion's position, we retrieved a separation of $34.4 \pm 0.1$~mas and a position angle of $16.8 \pm 0.3$~deg from the IFS data, so there is a slight discrepancy in the astrometry of the IFS and IRDIS data (see Table\,\ref{table:astrometry_photometry}). The difference in separation ($\approx$4~mas) could be related to the expected degeneracy with the contrast, which is somewhat higher at $K12$ compared to most of the IFS wavelength channels. The inferred position angle, on the other hand, is in agreement between the two instruments.

\subsection{Orbital configuration of HD\,142527\,B}
\label{sec:hd142527b_orbit}

\begin{figure*}
\centering
\includegraphics[width=0.8\linewidth]{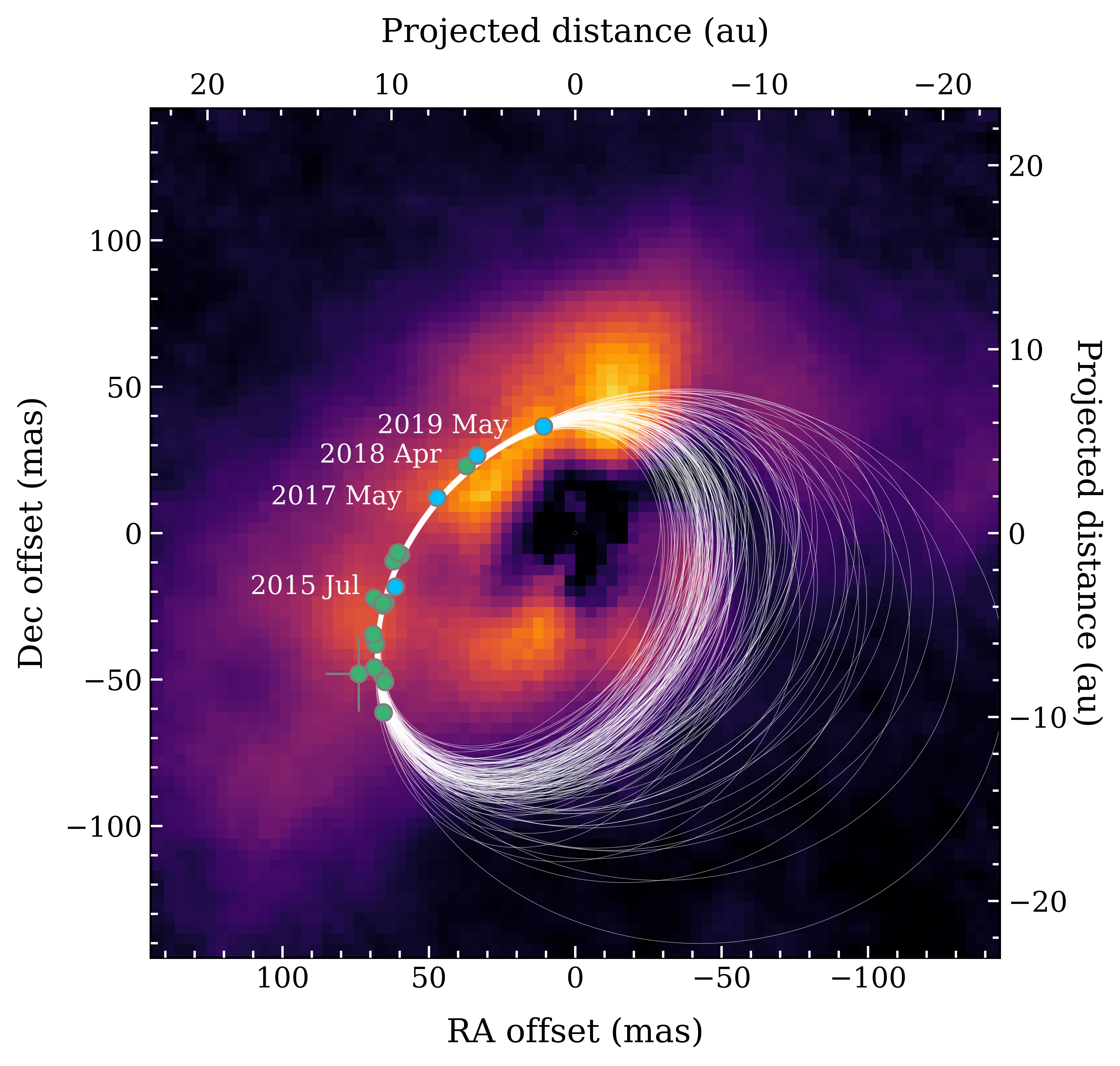}
\caption{Orbital analysis of HD\,142527\,B. The figure shows random samples from the posterior distribution, in comparison with the astrometric points that were used in the fit. The \emph{blue markers} are the astrometry from this work (see Table~\ref{table:astrometry_photometry}) and the \emph{green markers} are the values adopted from the literature. The orbital movement is in clockwise direction. The background shows the optical polarized scattered light image from \citet{avenhaus2017}.}
\label{fig:hd142527b_orbit}
\end{figure*}

The geometry of the HD\,142527 system is important for understanding its dynamical evolution and companion-disk interaction, in particular given the suggested misalignment between the inner and outer part of the circumstellar disk. The most recent orbital parameters were determined by \citet{balmer2022}, covering astrometric data from Mar 2012 to Apr 2018. Here, we will build forth on their work by complementing the available astrometry with the SPHERE/SAM point from May 2019, which is useful given the increasing curvature of the orbit at that epoch (see Table\,\ref{table:astrometry_photometry}). For completeness, we also include the SPHERE/SAM astrometry from the three reanalyzed epochs. However, as pointed out in Sect.\,\ref{sec:hd142527_astrometry_photometry}, these are consistent with the values from \citet{claudi2019} and therefore also the values used by \citet{balmer2022}.

We have estimated the orbital parameters by fitting the available astrometry from the literature \citep{biller2012,lacour2016,rodigas2014,christiaens2018,claudi2019,cugno2019,balmer2022} and the SPHERE/SAM astrometry from Table\,\ref{table:astrometry_photometry} with the \texttt{orbitize!}\footnote{\url{https://github.com/sblunt/orbitize}} package \citep{blunt2020}. The posterior distributions of the orbital elements were sampled with a parallel-tempered Markov chain Monte Carlo (MCMC) algorithm \citep{foreman2013,vousden2016}, while marginalizing over the parallax ($6.28 \pm 0.03$~mas; \citealt{gaia2016,gaia2023}) and system mass ($2.3 \pm 0.3$\,$\Msun$; \citealt{mendigutia2014,claudi2019}).

The parameters were estimated by using 20~temperatures, 200~walkers, and 50000~steps per walkers. We then removed the first 40000~steps as burn-in and selected every 20th step of each walker to exclude correlations between steps. Convergence of the chains was examined by calculating the integrated autocorrelation time and visually inspecting the trails of the walker's exploration of the posterior landscape. Figure~\ref{fig:hd142527b_orbit} shows 150 random orbit samples from the posterior distribution, in comparison with the inner dust structures that were detected in the polarized scattered light image by \citet{avenhaus2017} (see discussion in Sect.~\ref{sec:discussion_orbit}). From the posterior samples, we determined that the orbit has a semi-major axis of $a = 11.2^{+1.3}_{-0.6}$~au and an eccentricity of $e = 0.38^{+0.05}_{-0.05}$. The derived orbital period is $P = 26.4 \pm 7.1$~yr, and the periastron and apastron are $r_\mathrm{per} = 7.0^{+1.0}_{-0.8}$~au and $r_\mathrm{ap} = 15.3^{+2.1}_{-0.8}$~au. Figure~\ref{fig:posterior_orbit} in Appendix~\ref{sec:appendix_posteriors} displays the marginalized posterior distributions of all parameters.


The orientation of the circumstellar disk plane is defined by the inclination and position angle of the major axis. For the outer disk, we adopted $i_\mathrm{disk,out} = 38.2 \pm 1.3$~deg and $\Omega_\mathrm{disk,out} = 162.7 \pm 1.3$ from \citet{bohn2022} (although one of these two uncertainties is probably incorrectly printed in their Table~4 since all $i_\mathrm{out}$ and $\mathrm{PA}_\mathrm{out}$ pairs have the same uncertainty). With a velocity measurement, the latter value corresponds to the direction of redshifted line emission \citep[e.g.,][]{perez2015}, that is, the longitude of the ascending node, $\Omega_\mathrm{disk,out}$. If we assume that the spiral arms that are seen in scattered light \citep[e.g.,][]{avenhaus2014} have a trailing motion with respect to smaller separations, we can break the degeneracy on the inclination since this implies that the near side of the disk is in the western direction. From this we can conclude that the actual inclination of the outer disk is $i_\mathrm{disk,out} = 180 - 38.2 = 141.8 \pm 1.3$~deg when adopting the coordinate system as defined in \texttt{orbitize!}.

With the orbit fit, we constrained the orientation of the orbital plane of HD\,142527\,B to $i_\mathrm{orbit} = 139 \pm 4$~deg, and $\Omega_\mathrm{orbit} = 136 \pm 10$~deg or $316 \pm 10$~deg (correlated with the argument of periapsis, $\omega = 153 \pm 30$ and $\omega = -27 \pm 30$~deg, respectively). There is a 180~deg degeneracy without any radial velocity measurement of the companion so we can consider two cases for the mutual inclination between the companion's orbit at the outer disk plane. First, $\Omega_\mathrm{orbit} = 139 \pm 10$~deg would imply that the orbital plane of HD\,142527\,B is somewhat close to being coplanar with the midplane of the outer disk, since $i_\mathrm{orbit} = 139 \pm 4$~deg is consistent with $i_\mathrm{disk, out} = 141.8 \pm 1.3$~deg and $\Omega_\mathrm{orbit} = 136 \pm 10$~deg is comparable to $\Omega_\mathrm{disk,out} = 162.7 \pm 1.4$. The mutual inclination for this case is $\theta_\mathrm{out} = 18 \pm 6$~deg, which is calculated by propagating the posterior distribution from the orbit fit, while marginalizing over the uncertainty on the orientation of the disk plane. The second scenario is for $\Omega_\mathrm{orbit} = 316 \pm 10$~deg, which implies that the near side of the orbit is in NE direction. In that case, the mutual inclination is $\theta_\mathrm{out} = 76 \pm 4$~deg so the companion's orbit would be close to perpendicular to the outer disk.

For the inner disk, the situation is more complicated. The measurements by \citet{gravity_herbig_2019} provide an inclination, $\cos{i_\mathrm{disk,in}} = 0.91 \pm 0.02$ and position angle of the major axis $\mathrm{PA}_\mathrm{disk,in} = 14 \pm 4$~deg, but the near/far side is not known and there are not any velocity constraints. This implies that the actual inclination is either $\cos{i_\mathrm{disk,in}} = 0.91 \pm 0.02$ or $-0.91 \pm 0.02$, and the position angle of the ascending node is either $\Omega_\mathrm{disk,in} = 14 \pm 4$~deg or $194 \pm 4$~deg. So, there are four solutions for the mutual inclination between the companion's orbit and the inner disk: $\theta_\mathrm{in} = 34 \pm 6$~deg, $57 \pm 4$~deg, $123 \pm 4$~deg, or $146 \pm 6$~deg. Concluding, the orbit of HD\,142527\,B is significantly misaligned with respect to the inner disk for any of the solutions of the orbit-disk configuration.

\subsection{Atmospheric characterization of HD\,142527\,B}
\label{sec:hd142527b_atmosphere}

In this section, we will revisit the spectral characterization of HD\,142527\,B in the $YJHK$ bands with the Bayesian framework of \texttt{species} for estimating the atmospheric parameters. To do so, we fit the IFS $YJH$-band spectrum and the IRDIS photometry with a grid of BT-Settl model spectra \citep{allard2012}. We used the CIFIST release of the atmosphere model, which includes the updated solar abundances from \citet{caffau2011}. Apart from the atmospheric parameters ($T_\mathrm{eff}$, $\log{g}$), we fit simultaneously the visual extinction, $A_V$, using the empirical relation from \citet{cardelli1989} while fixing the reddening to $R_V = 3.1$. Furthermore, we included a Gaussian process to estimate the spectral covariances and we fit a relative error inflation to account for systematics that are not included in the error budget. The model fluxes are scaled to the data by fitting the radius, $R_\ast$, while using a Gaussian prior for the parallax, again from \emph{Gaia} DR3 \citep{gaia2016,gaia2023}. We also ran an additional fit that included a Gaussian prior for the mass of HD\,142527\,B, for which we adopted the dynamical measurement by \citet{claudi2019}, $M_\ast = 0.26 \pm 0.15$~$M_\odot$. The parameter estimation was done with \texttt{MultiNest} \citep{feroz2008,buchner2014}.

\begin{figure*}
\centering
\includegraphics[width=\linewidth]{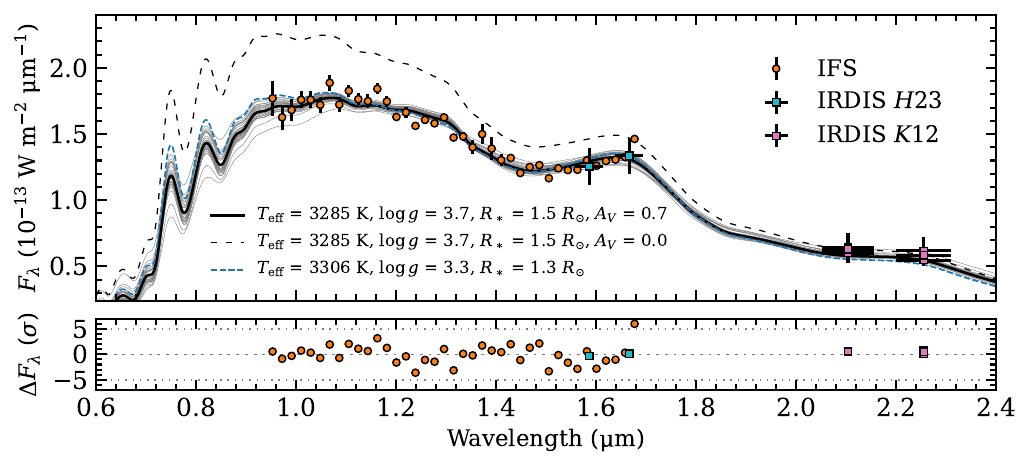}
\caption{Spectral analysis of HD\,142527\,B. The figure shows the best-fit model spectrum (\emph{black line}) at $R = 30$ in comparison with the IFS and IRDIS data (\emph{colored markers}). There are 30 randomly drawn spectra from the posterior distribution shown as \emph{gray lines}. The residuals (data minus model) are shown in the \emph{lower panel} relative to the data uncertainties. The \emph{black dashed line} is the best-fit spectrum that has been dereddened and the \emph{blue dashed line} is the best-fit without the extinction as free parameter.}
\label{fig:hd142527b_spectrum}
\end{figure*}

The best-fit spectrum and random posterior samples are shown in Fig.~\ref{fig:hd142527b_spectrum} in comparison with the data. The posterior distributions of the model parameters can be found in Fig.~\ref{fig:posterior_sed} of Appendix~\ref{sec:appendix_posteriors}. To summarize the main findings: we retrieved $T_\mathrm{eff} = 3285 \pm 49$~K, $\log{g} = 3.7 \pm 0.2$, $A_V = 0.7 \pm 0.3$, and $R_\ast = 1.46 \pm 0.08$~$R_\odot$. From the posterior samples, we calculated a bolometric luminosity of $\log\,L_\ast/L_\odot = -0.65 \pm 0.03$ and a companion mass of $\log\,M_\ast/M_\odot = -0.4 \pm 0.2$ (i.e., $M_\ast \approx 0.4$~$M_\odot$). The fractional amplitude of the correlated uncertainty peaks around 0.3, so about a third of the variances leak into off-axis terms of the covariance matrix, and the uncertainties of the IFS spectrum are inflated by 3\% (relative to the fluxes) to account for systematics that had not been captured with the spectral extraction and uncertainty estimation.

The residuals in Fig.~\ref{fig:hd142527b_spectrum}, which have been inflated with the retrieved uncertainty scaling, $b_\mathrm{SPHERE}$, seem reasonable at the $\approx$1--3$\sigma$ level. The residuals show marginally some systematic variations that could not be fit with the model spectra. We expect these variations to be caused by correlated noise in the data, instead of being actual features of astrophysical nature. This confirms the importance of including a Gaussian process to model the covariances in the SPHERE/IFS spectrum. Effectively, it reduces the overall weight in the likelihood function, thereby yielding more accurate posterior distributions. The correlation length that was retrieved with the Gaussian process is $\ell_\mathrm{SPHERE} \approx 0.01$~$\mu$m. This parameter provides the length scale by which the fit residuals would show a systematic variation that is caused by correlated noise, but it is not well constrained in this case (see Fig.~\ref{fig:posterior_sed}).

We also ran a fit with the mass prior, which acts on $\log\,g$. This led to a bit lower companion mass of $\log\,M_\ast/M_\odot = -0.5 \pm 0.2$, consistent with the prior mass, and a comparable radius of $R_\ast = 1.43 \pm 0.06$~$R_\odot$. Other retrieved parameters are also very similar to the case without the mass prior, so the difference in the spectral appearance is negligible. As a second test, we fit the SED without the extinction parameter, $A_V$. This yielded $T_\mathrm{eff} = 3306 \pm 40$~K and $R_\ast = 1.32 \pm 0.03$~$R_\odot$, but a slightly worse fit at several of the shortest and longest wavelength channels (see Fig.~\ref{fig:hd142527b_spectrum}). These findings were also confirmed by the Bayes' factor that is calculated from the marginalized likelihoods of the fit with and without extinction: $\Delta\ln{(Z)} = 2.2$. This suggests that, on the Jeffreys' scale \citep[e.g.,][]{trotta2008}, there is weak to moderate evidence that a model that includes $A_V$ as a free parameter is favored over a model without extinction.

The earlier mentioned bolometric luminosity, $\log{(L/L_\odot)} = -0.65 \pm 0.03$, was calculated by propagating the $T_\mathrm{eff}$ and $R_\ast$ samples from the posterior. This yielded the intrinsic luminosity, so assuming that photons are removed and not reemitted. From this luminosity, we estimated a companion mass of $M_\ast = 0.61 \pm 0.03$~$M_\odot$ and a radius of $R_\ast = 1.09 \pm 0.02$~$R_\odot$, based on the evolutionary tracks from \citet{baraffe2015} and adopting the age of $6.6^{+0.3}_{-1.5}$~Myr from \citet{vioque2018}, which is also consistent with the analysis by \citet{garufi2018}. This mass derived from the isochrone is a factor $\approx$2 higher (a $\approx$2$\sigma$ deviation) than the dynamical mass constraint from \citet{claudi2019}, $M_\ast = 0.26 \pm 0.15$~$M_\odot$, which the authors derived from the proper motion anomaly between multiple catalogs. The radius that is estimated from the evolutionary tracks is $\approx$0.4~$R_\odot$ smaller than the value estimated with the SED fit. By lowering the age to $2^{+2}_{-1}$~Myr \citep{fukagawa2006}, we determined $M_\ast = 0.39 \pm 0.11$\,$M_\odot$ and $R_\ast = 1.26 \pm 0.13$~$R_\odot$ from the evolution model, which are both consistent within the uncertainties with the inferred radius from the SED and the dynamical mass, respectively.

\section{Discussion}
\label{sec:discussion}

\subsection{Non-detections of companions in four transition disks}
\label{sec:discussion_limits}

The original classification of transition disks was based on a lack of IR emission seen in the SED indicating dust-depleted inner regions \citep{Espaillat2014}. The origin of the cavities is unclear, and while photoevaporation can be excluded for accreting objects \citep{Manara2014}, the presence of a massive companion, possibly stellar, was speculated early on. For example, using AMI observations, \citet{ireland2008} detected a binary star at 8~au from CoKu\,Tau/4 and discussed the possibility that other systems that had been classified as transition disks could also be circumbinary disks instead. 

As shown in Sect.~\ref{sec:chi2_map}, with our AMI observations we did not provide evidence for stellar or substellar companions within the cavities of the transition disks around HD\,100453, HD\,100546, HD\,135344\,B, and PDS\,70. For these systems, we can reject the presence of a (sub)stellar companion down to mass limits of approximately $60$, $40$, $80$, and $25$~$M_\mathrm{J}$ at separations of $\gtrsim$3--5~au. When adopting the stellar masses of our sample from \citet{vioque2018} and \citet{mueller2018}, we derived limits on the companion-to-star mass ratio, $q$, of a few percent ($q < 0.05$) for the four systems that we observed. For the archival data of HD\,142527, we derived a larger limit on the mass ratio of $q \approx 0.15$ and detected with high confidence the known companion. While the limits in terms of contrast are comparable, the high IR excess caused by the circumstellar disk leads to weaker limits in terms of mass.


We could rule out stellar companions down to $\approx$2~au in the four observed systems. The observations were not deep enough to reach the planetary-mass regime or even the regime of low- and intermediate-mass brown dwarfs, depending on the target (see Sect.~\ref{sec:sensitivity_limits}). The data would therefore still support a scenario with a multi-planet system or low-mass brown dwarf as the origin for the cavities. The multi-planet scenario has been explored in the theoretical work by \citet{dodsonrobinson2009} and \cite{Zhu2011}, and was later observationally confirmed with the case of PDS\,70. This star hosts two giant planets that orbit within the dust-depleted cavity \citep{keppler2018,haffert2019}, but our observations were not sensitive enough to reach the required contrast of $\approx$9~mag to detect PDS\,70\,b. Transition disks such as the one of PDS\,70 have been targeted by direct imaging surveys for over a decade now, since these are favorable locations to search for forming planets since effects by dust extinction are minimized \citep{sanchis2020}. Detection limits from coronagraphic imaging observations of HD\,100453, HD\,100546, and HD\,135344\,B are at best a few Jupiter masses, but only at large separations from the star, typically beyond the spatial extent of the disk \citep{asensio2021}.

Besides the direct detection of companions, the morphology of the substructures detected in the IR and (sub)millimeter are signposts of stellar and substellar companions, as well as giant planets. Such structures could in principle be used to discriminate between scenarios for the origin of cavities. For the two cases with known directly imaged low-mass companions, detailed hydrodynamical simulations have reproduced the observed disk features. As mentioned in Sect.~\ref{sec:introduction}, \citet{price2018} showed that HD\,142527\,B can be responsible for most of the substructures observed in HD\,142527, such as the cavity, horsehoe-shaped asymmetry, and the inner disk misalignment. For PDS\,70, simulations have shown that two giant planets in mean-motion resonance carve a common gap while migrating outward, with large dust being trapped at the pressure maxima in the outer disk \citep[e.g.,][]{bae2019,Toci2020}.

Binarity has been suspected in systems of intermediate-mass stars that host disks with a combined set of substructures: high NIR excess, spiral arms seen in scattered light, and shadows \citep{garufi2018}, and at millimeter wavelengths, (possibly eccentric) rings and azimuthal asymmetries \citep[e.g.,]{Boehler2018,Dong2018,Kraus2017}. An example of such objects is CQ\,Tau that additionally shows strong photometric variability at IR wavelengths \citep{Hammond2022}. All features - spirals, shadows and asymmetries - could be signatures of binaries and their detection supports the scenario that these disks may host yet-undetected stellar companions \citep{ragusa2017,ragusa2020,calcino2019,Rabago2023}. However, none of these stellar companions were found to date, and models including giant planets on eccentric orbits can also reproduce some of these features \citep{Baruteau2019,calcino2020,Pinilla2022}, so several scenarios remain viable.

More specifically for the observed targets, the origin of the spiral arms detected in HD\,100453, both at NIR \citep{benisty2017} and mm \citep{Rosotti2020} wavelengths, seems to be due to the widely separated, low-mass stellar companion \citep{Gonzalez2020}. The origin of the shadows, which point to a large misalignment between inner and outer disk \citep{min2017}, is still unclear. \citet{Zhu2019} found that massive planets on inclined orbits can break the inner disk from the outer disk, leading to clear observational signatures \citep{juhasz2017,Facchini2018}. In contrast, the origin of the spiral arms in the HD\,135344\,B remains debated, while the presence of shadows and their short term variability \citep{stolker2016b,stolker2017} indicate that the inner disk gets locally perturbed by some mechanism. Similarly, there is not much known about the origin of substructures that have been resolved in the disk around HD\,100546, even though the target is subject of candidate claims from various studies at a large range of wavelengths \citep[e.g.,][]{Brittain2014,Currie2015,Follette2017,Booth2023}.

While our observations could rule out the presence of stellar companions down to separations of a few astronomical units, the constraints on the presence of substellar companions are quite weak. A scenario in which a brown dwarf, or one or multiple giant planets are responsible for the observed substructures is therefore still worthwhile to further explore with observations. To do so, the \textit{James Webb Space Telescope} (\textit{JWST}) offers the opportunity to search for low-mass companions around the diffraction limit of the telescope with the AMI mode of the NIRISS instrument \citep{sivaramakrishnan2023}. At 3.5--5~$\mu$m, it could reach planetary masses if the closure phases can be sufficiently accurately calibrated \citep{kammerer2023}. At larger separations, the sensitivity by \textit{JWST} will place unique constraints on the presence of planetary-mass companion through coronagraphic imaging and NIR and MIR wavelengths. Searching for low-mass companions within transition disk cavities will therefore remain an exciting endeavor, requiring high-contrast imaging with space-based or AO-assisted observations.

\subsection{Toward deeper limits on companion masses}
\label{sec:deeper_limits}

The binary model that was used to search for companions and estimate detection limits was sufficient to reach into the brown dwarf regime and to extract the astrometry and (spectro)photometry of HD\,142527\,B with good precision for studying its orbit and atmosphere. To improve the sensitivity such that planetary masses can be reached there are two ways forward. First, the analysis is limited by systematics in the data that are known to be difficult to calibrate (see e.g., \citealt{lacour2011}). A better understanding of these systematic uncertainties may enable to include such effects in the forward model. Second, the model can be improved by including disk components that are known to be present in the data. All targets have a bright gap edge and/or resolved inner disk within the field of view of the AMI observations. Indeed, Fig.\,\ref{fig:chi2maps} shows such effects for several cases, HD\,100546 in particular.

The effect of a circumstellar disk on the closure phases is important to consider. A brightness asymmetry in a disk can mimic a companion, as has been debated for the case of T\,Cha \citep{huelamo2011,olofsson2013,cheetham2015a,sallum2015a}. A similar effect is seen in the detection map of HD\,100546 where two high-confidence features showed up on the near side of the gap edge. That side is brightest in total intensity due to forward scattering. We note again that the images in Fig\,\ref{fig:chi2maps} show polarized intensity so regions located along the major axis are typically brightest. At the smallest resolved separations, the observations may probe the outer regions on the inner disk components, as is the case for PDS\,70 and HD\,142527, and possibly also HD\,135344 (see Sect.\,\ref{sec:searching_companions}).

To robustly disentangle a potential companion and scattered light from the inner disk will require more detailed modeling that includes both a companion and disk emission. Such an approach is expected to yield deeper contrast limits, possibly also reaching into the planetary-mass regime for the Herbig stars of the sample. This was recently demonstrated by \citet{blakely2022} with SPHERE/SAM data on the LkCa\,15 system. They could infer the brightness distribution of both the gap edge that is also resolved with imaging in scattered light and the inner disk component that sits usually behind the coronagraph. Additionally, the simultaneous modeling excluded companions up to a contrast of $\approx$1000 within the gap where previously planets had been identified \citep{kraus2012,sallum2015b}. Inference on the presence of planets can be further complicated by temporal scattered light variations that can alter the position of an apparent point source. \citet{sallum2023} analyzed multi-epoch aperture masking observations of the LkCa\,15 system and showed that the data are consistent with a time-variable morphology. This points to either dynamical substructures at the forward-scattering side of the outer disk or variable shadowing by the inner disk. More detailed modeling of the closure phases from our observations, possibly combined with constraints from the SED and optical/NIR imaging, will therefore place quantitative constraints on the inner disk properties and deepen the detection limits. The impact of the systematics on the retrieved parameters will however become more severe at this level of precision. Forward modeling of such effects might be challenging, instead, it could also be investigated by testing with multiple observations and calibrators.

\subsection{Clumpy structures along the orbit of HD\,142527\,B}
\label{sec:discussion_orbit}

The orbital analysis in Sect.~\ref{sec:hd142527b_orbit} revealed an intriguing alignment of the orbit samples with the brightness distribution that is seen in the scattered light image. We also inferred that one of the two solutions for the mutual inclination is consistent with the orbit being coplanar with the outer disk. Here we will briefly discuss the constraints on the orbital elements, and we will speculate on the origin of the local disk features in the context of the companion's orbit.

With the AMI astrometry from the 2019 dataset, we have narrowed down the parameter space. Compared to the recent work by \citet{balmer2022}, all parameters are comparable at the $\approx$1--2$\sigma$ level, but the constraints have a higher precision because the 2019 epoch is at a phase where the curvature of the projected orbit is becoming stronger. With the fit, we inferred that the orbit has a semi-major axis of $a \approx 11$~au and an intermediate eccentricity of $e \approx 0.35$--$0.4$. Although the best-fit parameters are comparable to \citet{balmer2022}, we could reject their (low-probability) solutions that had a larger $a$ and $e$. We estimated that the companion passed its periastron in September 2020 ($\pm 100$~days), so significant orbital motion has taken place during the past few years years. 

The orbital constraints enable a comparison with the resolved disk features that are seen in scattered light. The polarized light image in Fig.~\ref{fig:hd142527b_orbit} was obtained by \citet{avenhaus2017} with SPHERE/ZIMPOL in the \emph{VBB} filter that covers the $R$ and $I$ bands. The extreme AO system of SPHERE had enabled an angular resolution of 34~mas at these optical wavelengths, so comparable to the $K$-band resolution that can be reached with SPHERE/SAM. The comparison of the image with the orbital constraints showed that the projected orbit appears aligned with the substructures seen in scattered light. We note that the scattered light traces dust at $\approx$10~au from the star, which is not necessarily the innermost disk detected with NIR interferometry \citep{gravity_herbig_2019} and likely associated with the strong NIR excess. \citet{avenhaus2017} suggested that the inner structures seen with ZIMPOL might be a halo-like component that extends to tens of au. Such a component had been previously introduced as the origin for the high NIR excess emission of this system, that can not be explained with a hydrostatically stable inner disk alone \citep{verhoeff2011}. We therefore interpret the resolved structures with SPHERE as an extended and diffuse component that is located beyond, or at least, at higher altitude than, the optically thick part of the inner disk.

It seems reasonable to assume that these disk features are substructures that originate from a dynamical interaction by HD\,142527\,B with its dusty circumstellar environment. The features appear unresolved, with the exception of the northern one that could be marginally resolved. Since we are seeing scattered light, brightness enhancements could trace a local increase of the density and/or temperature structure, but also a change in illumination geometry. Specifically, the scattered light flux will depend on the scattering angle. Since the azimuthal brightness variation is not continuous, we expect that an actual asymmetry in the structure of this extended component is causing the brightness variations instead of the phase function of the dust.

The scattered light image in Fig.~\ref{fig:hd142527b_orbit} shows the combined image of the 2015 and 2016 epochs by \citet{avenhaus2017}. At those epochs, the companion was located roughly at a position angle of $\sim$90~deg (i.e., in eastern direction), so it did not coincide with the projected location of the brightest disk feature. The individual images (see their Fig.~6) do show some slight differences that could be related to actual changes in the disk structure, as pointed out by \citet{avenhaus2017}. In Sect.~\ref{sec:hd142527b_orbit}, we estimated a period of $P = 26.4 \pm 7.1$~yr for the companion orbit so a one year difference is $\approx$4\% of a full orbit. Assuming that ejected dust clumps will settle toward their orbital plane on the dynamical timescale, it may seem reasonable to think that some changes in the brightness of the clumps are to be expected on a timescale of one year. Multi-epoch monitoring of both the disk features, in combination with astrometry and H$\alpha$ emission of the companion, would be required to investigate the companion-disk interaction and accretion processes in more detail.

As suggested, the clumpy structures in Fig.~\ref{fig:hd142527b_orbit} could potentially be the result of a dynamical perturbation of the small grains by the orbiting companion. The detection of hydrogen line emission indicates that the object is accreting \citep{close2014,cugno2019}, which happens at a variable rate \citep{balmer2022}. Therefore, HD\,142527\,B may possibly disturb its dusty environment and create substructures in the extended emission. Material is expected to be channeled from the circumstellar disk toward the companion atmosphere, and possibly mediated by a circumsecondary disk and a magnetosphere. This will cause a shock on the surface of the companion that could create an outburst by which material gets ejected into the surrounding. As a result, the clumpy structures could become more extended and optically thicker, which would enhance the amount of stellar light that gets scattered.

\subsection{Low-gravity atmosphere of HD\,142527\,B}
\label{sec:discussion_spectrum}

In Sect.~\ref{sec:hd142527b_atmosphere}, we modeled the SED of HD\,142527\,B with a grid of atmospheric spectra, by including the spectral and photometric measurements from this work and accounting for extinction. We found weak to moderate evidence that the SED is reddened by extinction and retrieved $A_V = 0.7 \pm 0.3$, which is consistent with estimates from the host star ($A_V \approx 0.6$; \citealt{verhoeff2011,garufi2018}). This may point to a common source of interstellar extinction, but we expected that there could also be a local component given the scattered light features that are present in the (projected) vicinity of the companion orbit. It seems therefore reasonable to conclude that some reddening is to be expected due to the extended and diffuse dusty structures from the inner disk. Similarly, the circumstellar environment of the primary star may also impact its extinction.

We derived $T_\mathrm{eff} = 3285 \pm 49$~K from the SED fit, which is a significantly higher value than the $T_\mathrm{eff} = 2600$--2800~K that was estimated by \citet{claudi2019}, also using BT-Settl model spectra. The difference might be in part explained by the extinction that was included in our fit, which allows for reddening and therefore opens the parameter space to higher $T_\mathrm{eff}$, although not including the extinction yielded a slightly higher $T_\mathrm{eff}$ of $3306 \pm 43$~K (see Fig.~\ref{fig:hd142527b_spectrum}). Perhaps more importantly, the IFS spectrum by \citet{claudi2019} that was obtained with the \emph{IRDIFS\_EXT} mode, appears by itself to have a somewhat redder slope (see their Fig.~3), which seems also the case for their IFS spectra obtained with AMI (see their Fig.~2). Potentially, this may point to a systematic difference with the spectral extraction and/or calibration spectrum that was used for the unresolved star and inner disk. Finally, the extinction may also vary along the companion's orbit, so the actual slope of the spectrum may have changed between epochs.

It is reassuring that without applying the mass prior, we could infer a spectroscopic companion mass that is consistent with the dynamical mass from \citet{claudi2019}. Related to the derived mass, we retrieved a surface gravity, $\log{g} = 3.7 \pm 0.2$, that seems physically in line with expectations for an $M$-type object at an age of several megayears. For the IFS wavelength range and resolution, the $H$ band is in particular an important diagnostic for the surface gravity, Specifically, the slight triangular shape, that is partly detected at the longest IFS wavelengths, helped to accurately constrain the surface gravity. We caution however that apart from $\log{g}$, also the metallicity will affect the relative depth of the absorption bands and therefore the $H$-band morphology. Specifically, a reduced metallicity will push the photosphere to higher pressures, thereby increasing the collision-induced absorption (CIA) and weakening the absorption bands. Since we retrieved a physical surface gravity and the model used solar abundances, we suggest that the composition of the atmosphere may have a solar metallicity since a nonsolar composition may have biased the surface gravity otherwise.

\citet{christiaens2018} carried out a detailed spectral and atmospheric characterization of HD\,142527\,B. The authors also used the BT-Setll model to infer the following parameters from their medium-resolution $H$- and $K$-band spectra: $T_\mathrm{eff} = 3500 \pm 100$~K, $\log{g} = 4.5 \pm 0.5$, and $A_H = 0.75 \pm 0.1$. The temperature is therefore consistent with our findings, but possibly a bit higher because also their retrieved extinction is higher (typically $A_V > A_H$). Between 1.5 and 1.7~$\mu$m their spectrum appears somewhat flatter (see their Fig.~5) which could in part be responsible for their higher surface gravity. Their derived radius, $R_\ast = 2.08 \pm 0.18$~$R_\odot$, is significantly higher, requiring either a sub-megayear age or additional emission from a circumsecondary environment \citep{christiaens2018}. The discrepancy in the retrieved radius seems to be mainly caused by the absolute fluxes which are a factor $\approx$1.3 higher in the $H$-band in the SINFONI spectrum. This presumably resulted in a larger radius since the adopted parallax is similar. The $H$-band flux in our IFS spectrum seems consistent with other spectral and photometric $H$-band measurements from the literature \citep{biller2012,lacour2016,claudi2019}. Discrepancies may point to systematic differences with the spectral extraction and/or stellar calibration. Or, there could be actual temporal variations in the spectral appearance since the companion is orbiting through a dusty environment.

\section{Conclusions}
\label{sec:conclusions}

We have reported on aperture masking observations with VLT/SPHERE of four young stars with a transition disk. The goal was to search for stellar and substellar companions that could be responsible for the central cavity and other disk features. We also analyzed four archival datasets of HD\,142527 to study its orbit and atmosphere. Here we summarize the main findings and conclusions:

\begin{itemize}
    \item The observations allowed us to explore the inner regions down to separations of $\approx$2~au where we can reject the presence of a stellar companion in all systems. At separations of $\gtrsim$3--5~au, we can reject substellar companions with lower mass limits in the range of 25--80~$M_\mathrm{J}$, depending on the target's age and brightness.

    \item In contrast to HD\,142527, the non-detections indicate that the cavities of HD\,100453, HD\,100546, HD\,135344\,B, and PDS\,70 are not shaped by the presence of a stellar companion, so we conclude that these systems are expected to indeed host a transition instead of circumbinary disk. Other mechanisms, such as the formation of a multi-planet system as the case of PDS\,70, are more likely to have carved the central regions.

    \item Dual-band imaging with SPHERE/SAM is a powerful technique to search for stellar and substellar companions around the diffraction limit of the telescope. AMI observations at longer wavelengths, such as $L$- and $M$-band imaging with the recently commissioned ERIS instrument, and the NIRISS instrument onboard \textit{JWST}, could enable detections of lower-mass companions, possibly within the planetary-mass regime.

    \item The orbital analysis of HD\,142527\,B revealed that the orbit is aligned with the projected disk structures that are seen in scattered light. This may point to a dynamical interaction between this low-mass companion and the extended structure of the inner disk. We also determined that the orbit could be close to coplanar with the outer disk and confirmed that it is misaligned with respect to the inner disk.

    \item The inferred atmospheric and bulk parameters from the (spectro)photometric data of HD\,142527\,B are in agreement with predictions by evolutionary models, pointing to a mass of $M_\ast = 0.4^{+0.3}_{-0.1}$~$M_\odot$ at an age of $6.6$~Myr. The spectroscopic mass is also consistent with dynamical constraints \citep{claudi2019}. The spectral appearance shows a reddening of $A_V = 0.7 \pm 0.3$, which is consistent with estimates of the host star ($A_V \approx 0.6$; \citealt{verhoeff2011,garufi2018}). Part of the extinction might be caused by the extended inner disk, both for the primary and secondary.

\end{itemize}

Aperture masking interferometry is an important technique for targeting low-mass companions around the diffraction limit at an intermediate contrast. High-precision astrometry from the \textit{Gaia} mission is expected to reveal a population of close-in substellar companions that will significantly increase the sample of known brown dwarf and planetary-mass companions. Aperture masking observations are going to play a valuable role with the atmospheric characterization of the brightest of such newly discovered companions.

\begin{acknowledgements}

The authors acknowledge the contribution of Attila Juhasz who initiated the science case and wrote the original proposal. The authors also want to thank William Balmer and Valentin Christiaens for helpful discussions and sharing of data. T.S.\ acknowledges the support from the Netherlands Organisation for Scientific Research (NWO) through grant VI.Veni.202.230. Part of this work was performed using the ALICE compute resources provided by Leiden University. This work used the Dutch national e-infrastructure with the support of the SURF Cooperative using grant no. EINF-1620. This project has received funding from the European Research Council (ERC) under the European Union's Horizon 2020 research and innovation programme (PROTOPLANETS, grant agreement No.~101002188). J.S.-B. acknowledges the support received from the UNAM PAPIIT project and IA 105023 and from the CONACyT “Ciencia de Frontera” project CF-2019/263975. This research has made use of the CDS Astronomical Databases SIMBAD and VIZIER.

\end{acknowledgements}

\bibliographystyle{aa}
\bibliography{references}

\begin{thebibliography}{130}
\expandafter\ifx\csname natexlab\endcsname\relax\def\natexlab#1{#1}\fi

\bibitem[{{Absil} {et~al.}(2011){Absil}, {Le Bouquin}, {Berger}, {Lagrange},
  {Chauvin}, {Lazareff}, {Zins}, {Haguenauer}, {Jocou}, {Kern}, {Millan-Gabet},
  {Rochat}, \& {Traub}}]{absil2011}
{Absil}, O., {Le Bouquin}, J.~B., {Berger}, J.~P., {et~al.} 2011, \aap, 535,
  A68

\bibitem[{{Allard} {et~al.}(2012){Allard}, {Homeier}, \&
  {Freytag}}]{allard2012}
{Allard}, F., {Homeier}, D., \& {Freytag}, B. 2012, Philosophical Transactions
  of the Royal Society of London Series A, 370, 2765

\bibitem[{{Andrews}(2020)}]{andrews2020}
{Andrews}, S.~M. 2020, \araa, 58, 483

\bibitem[{{Andrews} {et~al.}(2011){Andrews}, {Wilner}, {Espaillat}, {Hughes},
  {Dullemond}, {McClure}, {Qi}, \& {Brown}}]{andrews2011}
{Andrews}, S.~M., {Wilner}, D.~J., {Espaillat}, C., {et~al.} 2011, \apj, 732,
  42

\bibitem[{{Asensio-Torres} {et~al.}(2021){Asensio-Torres}, {Henning},
  {Cantalloube}, {Pinilla}, {Mesa}, {Garufi}, {Jorquera}, {Gratton}, {Chauvin},
  {Szul{\'a}gyi}, {van Boekel}, {Dong}, {Marleau}, {Benisty}, {Villenave},
  {Bergez-Casalou}, {Desgrange}, {Janson}, {Keppler}, {Langlois}, {M{\'e}nard},
  {Rickman}, {Stolker}, {Feldt}, {Fusco}, {Gluck}, {Pavlov}, \&
  {Ramos}}]{asensio2021}
{Asensio-Torres}, R., {Henning}, T., {Cantalloube}, F., {et~al.} 2021, \aap,
  652, A101

\bibitem[{{Avenhaus} {et~al.}(2017){Avenhaus}, {Quanz}, {Schmid}, {Dominik},
  {Stolker}, {Ginski}, {de Boer}, {Szul{\'a}gyi}, {Garufi}, {Zurlo},
  {Hagelberg}, {Benisty}, {Henning}, {M{\'e}nard}, {Meyer}, {Baruffolo},
  {Bazzon}, {Beuzit}, {Costille}, {Dohlen}, {Girard}, {Gisler}, {Kasper},
  {Mouillet}, {Pragt}, {Roelfsema}, {Salasnich}, \& {Sauvage}}]{avenhaus2017}
{Avenhaus}, H., {Quanz}, S.~P., {Schmid}, H.~M., {et~al.} 2017, \aj, 154, 33

\bibitem[{{Avenhaus} {et~al.}(2014){Avenhaus}, {Quanz}, {Schmid}, {Meyer},
  {Garufi}, {Wolf}, \& {Dominik}}]{avenhaus2014}
{Avenhaus}, H., {Quanz}, S.~P., {Schmid}, H.~M., {et~al.} 2014, \apj, 781, 87

\bibitem[{{Bae} {et~al.}(2019){Bae}, {Zhu}, {Baruteau}, {Benisty}, {Dullemond},
  {Facchini}, {Isella}, {Keppler}, {P{\'e}rez}, \& {Teague}}]{bae2019}
{Bae}, J., {Zhu}, Z., {Baruteau}, C., {et~al.} 2019, \apjl, 884, L41

\bibitem[{{Balmer} {et~al.}(2022){Balmer}, {Follette}, {Close}, {Males}, {De
  Rosa}, {Adams Redai}, {Watson}, {Weinberger}, {Morzinski}, {Morales},
  {Ward-Duong}, \& {Pueyo}}]{balmer2022}
{Balmer}, W.~O., {Follette}, K.~B., {Close}, L.~M., {et~al.} 2022, \aj, 164, 29

\bibitem[{{Baraffe} {et~al.}(2015){Baraffe}, {Homeier}, {Allard}, \&
  {Chabrier}}]{baraffe2015}
{Baraffe}, I., {Homeier}, D., {Allard}, F., \& {Chabrier}, G. 2015, Astronomy
  and Astrophysics, 577, A42

\bibitem[{{Baruteau} {et~al.}(2019){Baruteau}, {Barraza}, {P{\'e}rez},
  {Casassus}, {Dong}, {Lyra}, {Marino}, {Christiaens}, {Zhu}, {Carmona},
  {Debras}, \& {Alarcon}}]{Baruteau2019}
{Baruteau}, C., {Barraza}, M., {P{\'e}rez}, S., {et~al.} 2019, \mnras, 486, 304

\bibitem[{{Benisty} {et~al.}(2023){Benisty}, {Dominik}, {Follette}, {Garufi},
  {Ginski}, {Hashimoto}, {Keppler}, {Kley}, \& {Monnier}}]{benisty2023}
{Benisty}, M., {Dominik}, C., {Follette}, K., {et~al.} 2023, in Astronomical
  Society of the Pacific Conference Series, Vol. 534, Protostars and Planets
  VII, ed. S.~{Inutsuka}, Y.~{Aikawa}, T.~{Muto}, K.~{Tomida}, \& M.~{Tamura},
  605

\bibitem[{{Benisty} {et~al.}(2017){Benisty}, {Stolker}, {Pohl}, {de Boer},
  {Lesur}, {Dominik}, {Dullemond}, {Langlois}, {Min}, {Wagner}, {Henning},
  {Juhasz}, {Pinilla}, {Facchini}, {Apai}, {van Boekel}, {Garufi}, {Ginski},
  {M{\'e}nard}, {Pinte}, {Quanz}, {Zurlo}, {Boccaletti}, {Bonnefoy}, {Beuzit},
  {Chauvin}, {Cudel}, {Desidera}, {Feldt}, {Fontanive}, {Gratton}, {Kasper},
  {Lagrange}, {LeCoroller}, {Mouillet}, {Mesa}, {Sissa}, {Vigan}, {Antichi},
  {Buey}, {Fusco}, {Gisler}, {Llored}, {Magnard}, {Moeller-Nilsson}, {Pragt},
  {Roelfsema}, {Sauvage}, \& {Wildi}}]{benisty2017}
{Benisty}, M., {Stolker}, T., {Pohl}, A., {et~al.} 2017, \aap, 597, A42

\bibitem[{{Beuzit} {et~al.}(2019){Beuzit}, {Vigan}, {Mouillet}, {Dohlen},
  {Gratton}, {Boccaletti}, {Sauvage}, {Schmid}, {Langlois}, {Petit},
  {Baruffolo}, {Feldt}, {Milli}, {Wahhaj}, {Abe}, {Anselmi}, {Antichi},
  {Barette}, {Baudrand}, {Baudoz}, {Bazzon}, {Bernardi}, {Blanchard}, {Brast},
  {Bruno}, {Buey}, {Carbillet}, {Carle}, {Cascone}, {Chapron}, {Charton},
  {Chauvin}, {Claudi}, {Costille}, {De Caprio}, {de Boer}, {Delboulb{\'e}},
  {Desidera}, {Dominik}, {Downing}, {Dupuis}, {Fabron}, {Fantinel}, {Farisato},
  {Feautrier}, {Fedrigo}, {Fusco}, {Gigan}, {Ginski}, {Girard}, {Giro},
  {Gisler}, {Gluck}, {Gry}, {Henning}, {Hubin}, {Hugot}, {Incorvaia}, {Jaquet},
  {Kasper}, {Lagadec}, {Lagrange}, {Le Coroller}, {Le Mignant}, {Le Ruyet},
  {Lessio}, {Lizon}, {Llored}, {Lundin}, {Madec}, {Magnard}, {Marteaud},
  {Martinez}, {Maurel}, {M{\'e}nard}, {Mesa}, {M{\"o}ller-Nilsson}, {Moulin},
  {Moutou}, {Orign{\'e}}, {Parisot}, {Pavlov}, {Perret}, {Pragt}, {Puget},
  {Rabou}, {Ramos}, {Reess}, {Rigal}, {Rochat}, {Roelfsema}, {Rousset}, {Roux},
  {Saisse}, {Salasnich}, {Santambrogio}, {Scuderi}, {Segransan}, {Sevin},
  {Siebenmorgen}, {Soenke}, {Stadler}, {Suarez}, {Tiph{\`e}ne}, {Turatto},
  {Udry}, {Vakili}, {Waters}, {Weber}, {Wildi}, {Zins}, \&
  {Zurlo}}]{beuzit2019}
{Beuzit}, J.~L., {Vigan}, A., {Mouillet}, D., {et~al.} 2019, \aap, 631, A155

\bibitem[{{Biller} {et~al.}(2012){Biller}, {Lacour}, {Juh{\'a}sz}, {Benisty},
  {Chauvin}, {Olofsson}, {Pott}, {M{\"u}ller}, {Sicilia-Aguilar}, {Bonnefoy},
  {Tuthill}, {Thebault}, {Henning}, \& {Crida}}]{biller2012}
{Biller}, B., {Lacour}, S., {Juh{\'a}sz}, A., {et~al.} 2012, \apjl, 753, L38

\bibitem[{{Blakely} {et~al.}(2022){Blakely}, {Francis}, {Johnstone}, {Soulain},
  {Tuthill}, {Cheetham}, {Sanchez-Bermudez}, {Sivaramakrishnan}, {Dong}, {van
  der Marel}, {Cooper}, {Vigan}, \& {Cantalloube}}]{blakely2022}
{Blakely}, D., {Francis}, L., {Johnstone}, D., {et~al.} 2022, \apj, 931, 3

\bibitem[{{Blunt} {et~al.}(2020){Blunt}, {Wang}, {Angelo}, {Ngo}, {Cody}, {De
  Rosa}, {Graham}, {Hirsch}, {Nagpal}, {Nielsen}, {Pearce}, {Rice}, \&
  {Tejada}}]{blunt2020}
{Blunt}, S., {Wang}, J.~J., {Angelo}, I., {et~al.} 2020, \aj, 159, 89

\bibitem[{{Boehler} {et~al.}(2018){Boehler}, {Ricci}, {Weaver}, {Isella},
  {Benisty}, {Carpenter}, {Grady}, {Shen}, {Tang}, \& {Perez}}]{Boehler2018}
{Boehler}, Y., {Ricci}, L., {Weaver}, E., {et~al.} 2018, \apj, 853, 162

\bibitem[{{Bohlin} {et~al.}(2014){Bohlin}, {Gordon}, \&
  {Tremblay}}]{bohlin2014}
{Bohlin}, R.~C., {Gordon}, K.~D., \& {Tremblay}, P.~E. 2014, \pasp, 126, 711

\bibitem[{{Bohn} {et~al.}(2022){Bohn}, {Benisty}, {Perraut}, {van der Marel},
  {W{\"o}lfer}, {van Dishoeck}, {Facchini}, {Manara}, {Teague}, {Francis},
  {Berger}, {Garcia-Lopez}, {Ginski}, {Henning}, {Kenworthy}, {Kraus},
  {M{\'e}nard}, {M{\'e}rand}, \& {P{\'e}rez}}]{bohn2022}
{Bohn}, A.~J., {Benisty}, M., {Perraut}, K., {et~al.} 2022, \aap, 658, A183

\bibitem[{{Booth} {et~al.}(2023){Booth}, {Ilee}, {Walsh}, {Kama}, {Keyte}, {van
  Dishoeck}, \& {Nomura}}]{Booth2023}
{Booth}, A.~S., {Ilee}, J.~D., {Walsh}, C., {et~al.} 2023, \aap, 669, A53

\bibitem[{{Brittain} {et~al.}(2014){Brittain}, {Carr}, {Najita}, {Quanz}, \&
  {Meyer}}]{Brittain2014}
{Brittain}, S.~D., {Carr}, J.~S., {Najita}, J.~R., {Quanz}, S.~P., \& {Meyer},
  M.~R. 2014, \apj, 791, 136

\bibitem[{{Buchner} {et~al.}(2014){Buchner}, {Georgakakis}, {Nandra}, {Hsu},
  {Rangel}, {Brightman}, {Merloni}, {Salvato}, {Donley}, \&
  {Kocevski}}]{buchner2014}
{Buchner}, J., {Georgakakis}, A., {Nandra}, K., {et~al.} 2014, \aap, 564, A125

\bibitem[{{Caffau} {et~al.}(2011){Caffau}, {Ludwig}, {Steffen}, {Freytag}, \&
  {Bonifacio}}]{caffau2011}
{Caffau}, E., {Ludwig}, H.~G., {Steffen}, M., {Freytag}, B., \& {Bonifacio}, P.
  2011, \solphys, 268, 255

\bibitem[{{Calcino} {et~al.}(2020){Calcino}, {Christiaens}, {Price}, {Pinte},
  {Davis}, {van der Marel}, \& {Cuello}}]{calcino2020}
{Calcino}, J., {Christiaens}, V., {Price}, D.~J., {et~al.} 2020, \mnras, 498,
  639

\bibitem[{{Calcino} {et~al.}(2019){Calcino}, {Price}, {Pinte}, {van der Marel},
  {Ragusa}, {Dipierro}, {Cuello}, \& {Christiaens}}]{calcino2019}
{Calcino}, J., {Price}, D.~J., {Pinte}, C., {et~al.} 2019, \mnras, 490, 2579

\bibitem[{{Canovas} {et~al.}(2013){Canovas}, {M{\'e}nard}, {Hales},
  {Jord{\'a}n}, {Schreiber}, {Casassus}, {Gledhill}, \& {Pinte}}]{canovas2013}
{Canovas}, H., {M{\'e}nard}, F., {Hales}, A., {et~al.} 2013, \aap, 556, A123

\bibitem[{{Cardelli} {et~al.}(1989){Cardelli}, {Clayton}, \&
  {Mathis}}]{cardelli1989}
{Cardelli}, J.~A., {Clayton}, G.~C., \& {Mathis}, J.~S. 1989, \apj, 345, 245

\bibitem[{{Casassus} {et~al.}(2015){Casassus}, {Wright}, {Marino}, {Maddison},
  {Wootten}, {Roman}, {P{\'e}rez}, {Pinilla}, {Wyatt}, {Moral}, {M{\'e}nard},
  {Christiaens}, {Cieza}, \& {van der Plas}}]{casassus2015}
{Casassus}, S., {Wright}, C.~M., {Marino}, S., {et~al.} 2015, \apj, 812, 126

\bibitem[{{Cheetham} {et~al.}(2015{\natexlab{a}}){Cheetham}, {Huelamo},
  {Lacour}, {de Gregorio-Monsalvo}, \& {Tuthill}}]{cheetham2015a}
{Cheetham}, A., {Huelamo}, N., {Lacour}, S., {de Gregorio-Monsalvo}, I., \&
  {Tuthill}, P. 2015{\natexlab{a}}, \mnras, 450, L1

\bibitem[{{Cheetham} {et~al.}(2016){Cheetham}, {Girard}, {Lacour}, {Schworer},
  {Haubois}, \& {Beuzit}}]{cheetham2016}
{Cheetham}, A.~C., {Girard}, J., {Lacour}, S., {et~al.} 2016, in Society of
  Photo-Optical Instrumentation Engineers (SPIE) Conference Series, Vol. 9907,
  Optical and Infrared Interferometry and Imaging V, ed. F.~{Malbet}, M.~J.
  {Creech-Eakman}, \& P.~G. {Tuthill}, 99072T

\bibitem[{{Cheetham} {et~al.}(2015{\natexlab{b}}){Cheetham}, {Kraus},
  {Ireland}, {Cieza}, {Rizzuto}, \& {Tuthill}}]{cheetham2015b}
{Cheetham}, A.~C., {Kraus}, A.~L., {Ireland}, M.~J., {et~al.}
  2015{\natexlab{b}}, \apj, 813, 83

\bibitem[{{Cheetham} {et~al.}(2019){Cheetham}, {Samland}, {Brems}, {Launhardt},
  {Chauvin}, {S{\'e}gransan}, {Henning}, {Quirrenbach}, {Avenhaus}, {Cugno},
  {Girard}, {Godoy}, {Kennedy}, {Maire}, {Metchev}, {M{\"u}ller}, {Musso
  Barcucci}, {Olofsson}, {Pepe}, {Quanz}, {Queloz}, {Reffert}, {Rickman}, {van
  Boekel}, {Boccaletti}, {Bonnefoy}, {Cantalloube}, {Charnay}, {Delorme},
  {Janson}, {Keppler}, {Lagrange}, {Langlois}, {Lazzoni}, {Menard}, {Mesa},
  {Meyer}, {Schmidt}, {Sissa}, {Udry}, \& {Zurlo}}]{cheetham2019}
{Cheetham}, A.~C., {Samland}, M., {Brems}, S.~S., {et~al.} 2019, \aap, 622, A80

\bibitem[{{Chelli} {et~al.}(2016){Chelli}, {Duvert}, {Bourg{\`e}s}, {Mella},
  {Lafrasse}, {Bonneau}, \& {Chesneau}}]{Chelli2016}
{Chelli}, A., {Duvert}, G., {Bourg{\`e}s}, L., {et~al.} 2016, \aap, 589, A112

\bibitem[{{Christiaens} {et~al.}(2018){Christiaens}, {Casassus}, {Absil},
  {Kimeswenger}, {Gomez Gonzalez}, {Girard}, {Ram{\'\i}rez}, {Wertz}, {Zurlo},
  {Wahhaj}, {Flores}, {Salinas}, {Jord{\'a}n}, \& {Mawet}}]{christiaens2018}
{Christiaens}, V., {Casassus}, S., {Absil}, O., {et~al.} 2018, \aap, 617, A37

\bibitem[{{Christiaens} {et~al.}(2014){Christiaens}, {Casassus}, {Perez}, {van
  der Plas}, \& {M{\'e}nard}}]{christiaens2014}
{Christiaens}, V., {Casassus}, S., {Perez}, S., {van der Plas}, G., \&
  {M{\'e}nard}, F. 2014, \apjl, 785, L12

\bibitem[{{Claudi} {et~al.}(2019){Claudi}, {Maire}, {Mesa}, {Cheetham},
  {Fontanive}, {Gratton}, {Zurlo}, {Avenhaus}, {Bhowmik}, {Biller},
  {Boccaletti}, {Bonavita}, {Bonnefoy}, {Cascone}, {Chauvin}, {Delboulb{\'e}},
  {Desidera}, {D'Orazi}, {Feautrier}, {Feldt}, {Flammini Dotti}, {Girard},
  {Giro}, {Janson}, {Hagelberg}, {Keppler}, {Kopytova}, {Lacour}, {Lagrange},
  {Langlois}, {Lannier}, {Le Coroller}, {Menard}, {Messina}, {Meyer},
  {Millward}, {Olofsson}, {Pavlov}, {Peretti}, {Perrot}, {Pinte}, {Pragt},
  {Ramos}, {Rochat}, {Rodet}, {Roelfsema}, {Rouan}, {Salter}, {Schmidt},
  {Sissa}, {Thebault}, {Udry}, \& {Vigan}}]{claudi2019}
{Claudi}, R., {Maire}, A.~L., {Mesa}, D., {et~al.} 2019, \aap, 622, A96

\bibitem[{{Claudi} {et~al.}(2008){Claudi}, {Turatto}, {Gratton}, {Antichi},
  {Bonavita}, {Bruno}, {Cascone}, {De Caprio}, {Desidera}, {Giro}, {Mesa},
  {Scuderi}, {Dohlen}, {Beuzit}, \& {Puget}}]{claudi2008}
{Claudi}, R.~U., {Turatto}, M., {Gratton}, R.~G., {et~al.} 2008, in Society of
  Photo-Optical Instrumentation Engineers (SPIE) Conference Series, Vol. 7014,
  Ground-based and Airborne Instrumentation for Astronomy II, 70143E

\bibitem[{{Close} {et~al.}(2014){Close}, {Follette}, {Males}, {Puglisi},
  {Xompero}, {Apai}, {Najita}, {Weinberger}, {Morzinski}, {Rodigas}, {Hinz},
  {Bailey}, \& {Briguglio}}]{close2014}
{Close}, L.~M., {Follette}, K.~B., {Males}, J.~R., {et~al.} 2014, \apjl, 781,
  L30

\bibitem[{{Cugno} {et~al.}(2019){Cugno}, {Quanz}, {Hunziker}, {Stolker},
  {Schmid}, {Avenhaus}, {Baudoz}, {Bohn}, {Bonnefoy}, {Buenzli}, {Chauvin},
  {Cheetham}, {Desidera}, {Dominik}, {Feautrier}, {Feldt}, {Ginski}, {Girard},
  {Gratton}, {Hagelberg}, {Hugot}, {Janson}, {Lagrange}, {Langlois}, {Magnard},
  {Maire}, {Menard}, {Meyer}, {Milli}, {Mordasini}, {Pinte}, {Pragt},
  {Roelfsema}, {Rigal}, {Szul{\'a}gyi}, {van Boekel}, {van der Plas}, {Vigan},
  {Wahhaj}, \& {Zurlo}}]{cugno2019}
{Cugno}, G., {Quanz}, S.~P., {Hunziker}, S., {et~al.} 2019, \aap, 622, A156

\bibitem[{{Currie} {et~al.}(2015){Currie}, {Cloutier}, {Brittain}, {Grady},
  {Burrows}, {Muto}, {Kenyon}, \& {Kuchner}}]{Currie2015}
{Currie}, T., {Cloutier}, R., {Brittain}, S., {et~al.} 2015, \apjl, 814, L27

\bibitem[{{Cushing} {et~al.}(2004){Cushing}, {Vacca}, \&
  {Rayner}}]{cushing2004}
{Cushing}, M.~C., {Vacca}, W.~D., \& {Rayner}, J.~T. 2004, \pasp, 116, 362

\bibitem[{{Cutri} {et~al.}(2003){Cutri}, {Skrutskie}, {van Dyk}, {Beichman},
  {Carpenter}, {Chester}, {Cambresy}, {Evans}, {Fowler}, {Gizis}, {Howard},
  {Huchra}, {Jarrett}, {Kopan}, {Kirkpatrick}, {Light}, {Marsh}, {McCallon},
  {Schneider}, {Stiening}, {Sykes}, {Weinberg}, {Wheaton}, {Wheelock}, \&
  {Zacarias}}]{cutri2003}
{Cutri}, R.~M., {Skrutskie}, M.~F., {van Dyk}, S., {et~al.} 2003, {2MASS All
  Sky Catalog of point sources.}

\bibitem[{{Dodson-Robinson} {et~al.}(2009){Dodson-Robinson}, {Veras}, {Ford},
  \& {Beichman}}]{dodsonrobinson2009}
{Dodson-Robinson}, S.~E., {Veras}, D., {Ford}, E.~B., \& {Beichman}, C.~A.
  2009, \apj, 707, 79

\bibitem[{{Dohlen} {et~al.}(2008){Dohlen}, {Langlois}, {Saisse}, {Hill},
  {Origne}, {Jacquet}, {Fabron}, {Blanc}, {Llored}, {Carle}, {Moutou}, {Vigan},
  {Boccaletti}, {Carbillet}, {Mouillet}, \& {Beuzit}}]{dohlen2008}
{Dohlen}, K., {Langlois}, M., {Saisse}, M., {et~al.} 2008, in Society of
  Photo-Optical Instrumentation Engineers (SPIE) Conference Series, Vol. 7014,
  \procspie, 70143L

\bibitem[{{Dong} {et~al.}(2018){Dong}, {Liu}, {Eisner}, {Andrews}, {Fung},
  {Zhu}, {Chiang}, {Hashimoto}, {Liu}, {Casassus}, {Esposito}, {Hasegawa},
  {Muto}, {Pavlyuchenkov}, {Wilner}, {Akiyama}, {Tamura}, \&
  {Wisniewski}}]{Dong2018}
{Dong}, R., {Liu}, S.-y., {Eisner}, J., {et~al.} 2018, \apj, 860, 124

\bibitem[{{Espaillat} {et~al.}(2014){Espaillat}, {Muzerolle}, {Najita},
  {Andrews}, {Zhu}, {Calvet}, {Kraus}, {Hashimoto}, {Kraus}, \&
  {D'Alessio}}]{Espaillat2014}
{Espaillat}, C., {Muzerolle}, J., {Najita}, J., {et~al.} 2014, in Protostars
  and Planets VI, ed. H.~{Beuther}, R.~S. {Klessen}, C.~P. {Dullemond}, \&
  T.~{Henning}, 497--520

\bibitem[{{Evans} {et~al.}(2012){Evans}, {Ireland}, {Kraus}, {Martinache},
  {Stewart}, {Tuthill}, {Lacour}, {Carpenter}, \& {Hillenbrand}}]{evans2012}
{Evans}, T.~M., {Ireland}, M.~J., {Kraus}, A.~L., {et~al.} 2012, \apj, 744, 120

\bibitem[{{Facchini} {et~al.}(2018){Facchini}, {Juh{\'a}sz}, \&
  {Lodato}}]{Facchini2018}
{Facchini}, S., {Juh{\'a}sz}, A., \& {Lodato}, G. 2018, \mnras, 473, 4459

\bibitem[{{Feroz} \& {Hobson}(2008)}]{feroz2008}
{Feroz}, F. \& {Hobson}, M.~P. 2008, \mnras, 384, 449

\bibitem[{{Follette} {et~al.}(2017){Follette}, {Rameau}, {Dong}, {Pueyo},
  {Close}, {Duch{\^e}ne}, {Fung}, {Leonard}, {Macintosh}, {Males}, {Marois},
  {Millar-Blanchaer}, {Morzinski}, {Mullen}, {Perrin}, {Spiro}, {Wang},
  {Ammons}, {Bailey}, {Barman}, {Bulger}, {Chilcote}, {Cotten}, {De Rosa},
  {Doyon}, {Fitzgerald}, {Goodsell}, {Graham}, {Greenbaum}, {Hibon}, {Hung},
  {Ingraham}, {Kalas}, {Konopacky}, {Larkin}, {Maire}, {Marchis}, {Metchev},
  {Nielsen}, {Oppenheimer}, {Palmer}, {Patience}, {Poyneer}, {Rajan},
  {Rantakyr{\"o}}, {Savransky}, {Schneider}, {Sivaramakrishnan}, {Song},
  {Soummer}, {Thomas}, {Vega}, {Wallace}, {Ward-Duong}, {Wiktorowicz}, \&
  {Wolff}}]{Follette2017}
{Follette}, K.~B., {Rameau}, J., {Dong}, R., {et~al.} 2017, \aj, 153, 264

\bibitem[{{Foreman-Mackey} {et~al.}(2013){Foreman-Mackey}, {Hogg}, {Lang}, \&
  {Goodman}}]{foreman2013}
{Foreman-Mackey}, D., {Hogg}, D.~W., {Lang}, D., \& {Goodman}, J. 2013, \pasp,
  125, 306

\bibitem[{{Francis} \& {van der Marel}(2020)}]{francis2020}
{Francis}, L. \& {van der Marel}, N. 2020, \apj, 892, 111

\bibitem[{{Fukagawa} {et~al.}(2006){Fukagawa}, {Tamura}, {Itoh}, {Kudo},
  {Imaeda}, {Oasa}, {Hayashi}, \& {Hayashi}}]{fukagawa2006}
{Fukagawa}, M., {Tamura}, M., {Itoh}, Y., {et~al.} 2006, \apjl, 636, L153

\bibitem[{{Fukagawa} {et~al.}(2013){Fukagawa}, {Tsukagoshi}, {Momose}, {Saigo},
  {Ohashi}, {Kitamura}, {Inutsuka}, {Muto}, {Nomura}, {Takeuchi}, {Kobayashi},
  {Hanawa}, {Akiyama}, {Honda}, {Fujiwara}, {Kataoka}, {Takahashi}, \&
  {Shibai}}]{fukagawa2013}
{Fukagawa}, M., {Tsukagoshi}, T., {Momose}, M., {et~al.} 2013, \pasj, 65, L14

\bibitem[{{Fusco} {et~al.}(2006){Fusco}, {Rousset}, {Sauvage}, {Petit},
  {Beuzit}, {Dohlen}, {Mouillet}, {Charton}, {Nicolle}, {Kasper}, {Baudoz}, \&
  {Puget}}]{fusco2006}
{Fusco}, T., {Rousset}, G., {Sauvage}, J.~F., {et~al.} 2006, Optics Express,
  14, 7515

\bibitem[{{Gaia Collaboration} {et~al.}(2016){Gaia Collaboration}, {Prusti},
  {de Bruijne}, {Brown}, {Vallenari}, {Babusiaux}, {Bailer-Jones}, {Bastian},
  {Biermann}, {Evans}, {Eyer}, {Jansen}, {Jordi}, {Klioner}, {Lammers},
  {Lindegren}, {Luri}, {Mignard}, {Milligan}, {Panem}, {Poinsignon},
  {Pourbaix}, {Randich}, {Sarri}, {Sartoretti}, {Siddiqui}, {Soubiran},
  {Valette}, {van Leeuwen}, {Walton}, {Aerts}, {Arenou}, {Cropper}, {Drimmel},
  {H{\o}g}, {Katz}, {Lattanzi}, {O'Mullane}, {Grebel}, {Holland}, {Huc},
  {Passot}, {Bramante}, {Cacciari}, {Casta{\~n}eda}, {Chaoul}, {Cheek}, {De
  Angeli}, {Fabricius}, {Guerra}, {Hern{\'a}ndez}, {Jean-Antoine-Piccolo},
  {Masana}, {Messineo}, {Mowlavi}, {Nienartowicz}, {Ord{\'o}{\~n}ez-Blanco},
  {Panuzzo}, {Portell}, {Richards}, {Riello}, {Seabroke}, {Tanga},
  {Th{\'e}venin}, {Torra}, {Els}, {Gracia-Abril}, {Comoretto},
  {Garcia-Reinaldos}, {Lock}, {Mercier}, {Altmann}, {Andrae}, {Astraatmadja},
  {Bellas-Velidis}, {Benson}, {Berthier}, {Blomme}, {Busso}, {Carry},
  {Cellino}, {Clementini}, {Cowell}, {Creevey}, {Cuypers}, {Davidson}, {De
  Ridder}, {de Torres}, {Delchambre}, {Dell'Oro}, {Ducourant}, {Fr{\'e}mat},
  {Garc{\'\i}a-Torres}, {Gosset}, {Halbwachs}, {Hambly}, {Harrison}, {Hauser},
  {Hestroffer}, {Hodgkin}, {Huckle}, {Hutton}, {Jasniewicz}, {Jordan},
  {Kontizas}, {Korn}, {Lanzafame}, {Manteiga}, {Moitinho}, {Muinonen},
  {Osinde}, {Pancino}, {Pauwels}, {Petit}, {Recio-Blanco}, {Robin}, {Sarro},
  {Siopis}, {Smith}, {Smith}, {Sozzetti}, {Thuillot}, {van Reeven}, {Viala},
  {Abbas}, {Abreu Aramburu}, {Accart}, {Aguado}, {Allan}, {Allasia},
  {Altavilla}, {{\'A}lvarez}, {Alves}, {Anderson}, {Andrei}, {Anglada Varela},
  {Antiche}, {Antoja}, {Ant{\'o}n}, {Arcay}, {Atzei}, {Ayache}, {Bach},
  {Baker}, {Balaguer-N{\'u}{\~n}ez}, {Barache}, {Barata}, {Barbier}, {Barblan},
  {Baroni}, {Barrado y Navascu{\'e}s}, {Barros}, {Barstow}, {Becciani},
  {Bellazzini}, {Bellei}, {Bello Garc{\'\i}a}, {Belokurov}, {Bendjoya},
  {Berihuete}, {Bianchi}, {Bienaym{\'e}}, {Billebaud}, {Blagorodnova},
  {Blanco-Cuaresma}, {Boch}, {Bombrun}, {Borrachero}, {Bouquillon}, {Bourda},
  {Bouy}, {Bragaglia}, {Breddels}, {Brouillet}, {Br{\"u}semeister},
  {Bucciarelli}, {Budnik}, {Burgess}, {Burgon}, {Burlacu}, {Busonero}, {Buzzi},
  {Caffau}, {Cambras}, {Campbell}, {Cancelliere}, {Cantat-Gaudin}, {Carlucci},
  {Carrasco}, {Castellani}, {Charlot}, {Charnas}, {Charvet}, {Chassat},
  {Chiavassa}, {Clotet}, {Cocozza}, {Collins}, {Collins}, {Costigan}, {Crifo},
  {Cross}, {Crosta}, {Crowley}, {Dafonte}, {Damerdji}, {Dapergolas}, {David},
  {David}, {De Cat}, {de Felice}, {de Laverny}, {De Luise}, {De March}, {de
  Martino}, {de Souza}, {Debosscher}, {del Pozo}, {Delbo}, {Delgado},
  {Delgado}, {di Marco}, {Di Matteo}, {Diakite}, {Distefano}, {Dolding}, {Dos
  Anjos}, {Drazinos}, {Dur{\'a}n}, {Dzigan}, {Ecale}, {Edvardsson}, {Enke},
  {Erdmann}, {Escolar}, {Espina}, {Evans}, {Eynard Bontemps}, {Fabre},
  {Fabrizio}, {Faigler}, {Falc{\~a}o}, {Farr{\`a}s Casas}, {Faye}, {Federici},
  {Fedorets}, {Fern{\'a}ndez-Hern{\'a}ndez}, {Fernique}, {Fienga}, {Figueras},
  {Filippi}, {Findeisen}, {Fonti}, {Fouesneau}, {Fraile}, {Fraser}, {Fuchs},
  {Furnell}, {Gai}, {Galleti}, {Galluccio}, {Garabato}, {Garc{\'\i}a-Sedano},
  {Gar{\'e}}, {Garofalo}, {Garralda}, {Gavras}, {Gerssen}, {Geyer}, {Gilmore},
  {Girona}, {Giuffrida}, {Gomes}, {Gonz{\'a}lez-Marcos},
  {Gonz{\'a}lez-N{\'u}{\~n}ez}, {Gonz{\'a}lez-Vidal}, {Granvik}, {Guerrier},
  {Guillout}, {Guiraud}, {G{\'u}rpide}, {Guti{\'e}rrez-S{\'a}nchez}, {Guy},
  {Haigron}, {Hatzidimitriou}, {Haywood}, {Heiter}, {Helmi}, {Hobbs},
  {Hofmann}, {Holl}, {Holland}, {Hunt}, {Hypki}, {Icardi}, {Irwin}, {Jevardat
  de Fombelle}, {Jofr{\'e}}, {Jonker}, {Jorissen}, {Julbe}, {Karampelas},
  {Kochoska}, {Kohley}, {Kolenberg}, {Kontizas}, {Koposov}, {Kordopatis},
  {Koubsky}, {Kowalczyk}, {Krone-Martins}, {Kudryashova}, {Kull}, {Bachchan},
  {Lacoste-Seris}, {Lanza}, {Lavigne}, {Le Poncin-Lafitte}, {Lebreton},
  {Lebzelter}, {Leccia}, {Leclerc}, {Lecoeur-Taibi}, {Lemaitre}, {Lenhardt},
  {Leroux}, {Liao}, {Licata}, {Lindstr{\o}m}, {Lister}, {Livanou}, {Lobel},
  {L{\"o}ffler}, {L{\'o}pez}, {Lopez-Lozano}, {Lorenz}, {Loureiro},
  {MacDonald}, {Magalh{\~a}es Fernandes}, {Managau}, {Mann}, {Mantelet},
  {Marchal}, {Marchant}, {Marconi}, {Marie}, {Marinoni}, {Marrese},
  {Marschalk{\'o}}, {Marshall}, {Mart{\'\i}n-Fleitas}, {Martino}, {Mary},
  {Matijevi{\v{c}}}, {Mazeh}, {McMillan}, {Messina}, {Mestre}, {Michalik},
  {Millar}, {Miranda}, {Molina}, {Molinaro}, {Molinaro}, {Moln{\'a}r},
  {Moniez}, {Montegriffo}, {Monteiro}, {Mor}, {Mora}, {Morbidelli}, {Morel},
  {Morgenthaler}, {Morley}, {Morris}, {Mulone}, {Muraveva}, {Musella},
  {Narbonne}, {Nelemans}, {Nicastro}, {Noval}, {Ord{\'e}novic},
  {Ordieres-Mer{\'e}}, {Osborne}, {Pagani}, {Pagano}, {Pailler}, {Palacin},
  {Palaversa}, {Parsons}, {Paulsen}, {Pecoraro}, {Pedrosa}, {Pentik{\"a}inen},
  {Pereira}, {Pichon}, {Piersimoni}, {Pineau}, {Plachy}, {Plum}, {Poujoulet},
  {Pr{\v{s}}a}, {Pulone}, {Ragaini}, {Rago}, {Rambaux}, {Ramos-Lerate},
  {Ranalli}, {Rauw}, {Read}, {Regibo}, {Renk}, {Reyl{\'e}}, {Ribeiro},
  {Rimoldini}, {Ripepi}, {Riva}, {Rixon}, {Roelens}, {Romero-G{\'o}mez},
  {Rowell}, {Royer}, {Rudolph}, {Ruiz-Dern}, {Sadowski}, {Sagrist{\`a}
  Sell{\'e}s}, {Sahlmann}, {Salgado}, {Salguero}, {Sarasso}, {Savietto},
  {Schnorhk}, {Schultheis}, {Sciacca}, {Segol}, {Segovia}, {Segransan},
  {Serpell}, {Shih}, {Smareglia}, {Smart}, {Smith}, {Solano}, {Solitro},
  {Sordo}, {Soria Nieto}, {Souchay}, {Spagna}, {Spoto}, {Stampa}, {Steele},
  {Steidelm{\"u}ller}, {Stephenson}, {Stoev}, {Suess}, {S{\"u}veges}, {Surdej},
  {Szabados}, {Szegedi-Elek}, {Tapiador}, {Taris}, {Tauran}, {Taylor},
  {Teixeira}, {Terrett}, {Tingley}, {Trager}, {Turon}, {Ulla}, {Utrilla},
  {Valentini}, {van Elteren}, {Van Hemelryck}, {van Leeuwen}, {Varadi},
  {Vecchiato}, {Veljanoski}, {Via}, {Vicente}, {Vogt}, {Voss}, {Votruba},
  {Voutsinas}, {Walmsley}, {Weiler}, {Weingrill}, {Werner}, {Wevers},
  {Whitehead}, {Wyrzykowski}, {Yoldas}, {{\v{Z}}erjal}, {Zucker}, {Zurbach},
  {Zwitter}, {Alecu}, {Allen}, {Allende Prieto}, {Amorim},
  {Anglada-Escud{\'e}}, {Arsenijevic}, {Azaz}, {Balm}, {Beck}, {Bernstein},
  {Bigot}, {Bijaoui}, {Blasco}, {Bonfigli}, {Bono}, {Boudreault}, {Bressan},
  {Brown}, {Brunet}, {Bunclark}, {Buonanno}, {Butkevich}, {Carret}, {Carrion},
  {Chemin}, {Ch{\'e}reau}, {Corcione}, {Darmigny}, {de Boer}, {de Teodoro}, {de
  Zeeuw}, {Delle Luche}, {Domingues}, {Dubath}, {Fodor}, {Fr{\'e}zouls},
  {Fries}, {Fustes}, {Fyfe}, {Gallardo}, {Gallegos}, {Gardiol}, {Gebran},
  {Gomboc}, {G{\'o}mez}, {Grux}, {Gueguen}, {Heyrovsky}, {Hoar}, {Iannicola},
  {Isasi Parache}, {Janotto}, {Joliet}, {Jonckheere}, {Keil}, {Kim},
  {Klagyivik}, {Klar}, {Knude}, {Kochukhov}, {Kolka}, {Kos}, {Kutka}, {Lainey},
  {LeBouquin}, {Liu}, {Loreggia}, {Makarov}, {Marseille}, {Martayan},
  {Martinez-Rubi}, {Massart}, {Meynadier}, {Mignot}, {Munari}, {Nguyen},
  {Nordlander}, {Ocvirk}, {O'Flaherty}, {Olias Sanz}, {Ortiz}, {Osorio},
  {Oszkiewicz}, {Ouzounis}, {Palmer}, {Park}, {Pasquato}, {Peltzer}, {Peralta},
  {P{\'e}turaud}, {Pieniluoma}, {Pigozzi}, {Poels}, {Prat}, {Prod'homme},
  {Raison}, {Rebordao}, {Risquez}, {Rocca-Volmerange}, {Rosen}, {Ruiz-Fuertes},
  {Russo}, {Sembay}, {Serraller Vizcaino}, {Short}, {Siebert}, {Silva},
  {Sinachopoulos}, {Slezak}, {Soffel}, {Sosnowska}, {Strai{\v{z}}ys}, {ter
  Linden}, {Terrell}, {Theil}, {Tiede}, {Troisi}, {Tsalmantza}, {Tur},
  {Vaccari}, {Vachier}, {Valles}, {Van Hamme}, {Veltz}, {Virtanen}, {Wallut},
  {Wichmann}, {Wilkinson}, {Ziaeepour}, \& {Zschocke}}]{gaia2016}
{Gaia Collaboration}, {Prusti}, T., {de Bruijne}, J.~H.~J., {et~al.} 2016,
  \aap, 595, A1

\bibitem[{{Gaia Collaboration} {et~al.}(2023){Gaia Collaboration}, {Vallenari},
  {Brown}, {Prusti}, {de Bruijne}, {Arenou}, {Babusiaux}, {Biermann},
  {Creevey}, {Ducourant}, {Evans}, {Eyer}, {Guerra}, {Hutton}, {Jordi},
  {Klioner}, {Lammers}, {Lindegren}, {Luri}, {Mignard}, {Panem}, {Pourbaix},
  {Randich}, {Sartoretti}, {Soubiran}, {Tanga}, {Walton}, {Bailer-Jones},
  {Bastian}, {Drimmel}, {Jansen}, {Katz}, {Lattanzi}, {van Leeuwen}, {Bakker},
  {Cacciari}, {Casta{\~n}eda}, {De Angeli}, {Fabricius}, {Fouesneau},
  {Fr{\'e}mat}, {Galluccio}, {Guerrier}, {Heiter}, {Masana}, {Messineo},
  {Mowlavi}, {Nicolas}, {Nienartowicz}, {Pailler}, {Panuzzo}, {Riclet}, {Roux},
  {Seabroke}, {Sordo}, {Th{\'e}venin}, {Gracia-Abril}, {Portell}, {Teyssier},
  {Altmann}, {Andrae}, {Audard}, {Bellas-Velidis}, {Benson}, {Berthier},
  {Blomme}, {Burgess}, {Busonero}, {Busso}, {C{\'a}novas}, {Carry}, {Cellino},
  {Cheek}, {Clementini}, {Damerdji}, {Davidson}, {de Teodoro}, {Nu{\~n}ez
  Campos}, {Delchambre}, {Dell'Oro}, {Esquej}, {Fern{\'a}ndez-Hern{\'a}ndez},
  {Fraile}, {Garabato}, {Garc{\'\i}a-Lario}, {Gosset}, {Haigron}, {Halbwachs},
  {Hambly}, {Harrison}, {Hern{\'a}ndez}, {Hestroffer}, {Hodgkin}, {Holl},
  {Jan{\ss}en}, {Jevardat de Fombelle}, {Jordan}, {Krone-Martins}, {Lanzafame},
  {L{\"o}ffler}, {Marchal}, {Marrese}, {Moitinho}, {Muinonen}, {Osborne},
  {Pancino}, {Pauwels}, {Recio-Blanco}, {Reyl{\'e}}, {Riello}, {Rimoldini},
  {Roegiers}, {Rybizki}, {Sarro}, {Siopis}, {Smith}, {Sozzetti}, {Utrilla},
  {van Leeuwen}, {Abbas}, {{\'A}brah{\'a}m}, {Abreu Aramburu}, {Aerts},
  {Aguado}, {Ajaj}, {Aldea-Montero}, {Altavilla}, {{\'A}lvarez}, {Alves},
  {Anders}, {Anderson}, {Anglada Varela}, {Antoja}, {Baines}, {Baker},
  {Balaguer-N{\'u}{\~n}ez}, {Balbinot}, {Balog}, {Barache}, {Barbato},
  {Barros}, {Barstow}, {Bartolom{\'e}}, {Bassilana}, {Bauchet}, {Becciani},
  {Bellazzini}, {Berihuete}, {Bernet}, {Bertone}, {Bianchi}, {Binnenfeld},
  {Blanco-Cuaresma}, {Blazere}, {Boch}, {Bombrun}, {Bossini}, {Bouquillon},
  {Bragaglia}, {Bramante}, {Breedt}, {Bressan}, {Brouillet}, {Brugaletta},
  {Bucciarelli}, {Burlacu}, {Butkevich}, {Buzzi}, {Caffau}, {Cancelliere},
  {Cantat-Gaudin}, {Carballo}, {Carlucci}, {Carnerero}, {Carrasco},
  {Casamiquela}, {Castellani}, {Castro-Ginard}, {Chaoul}, {Charlot}, {Chemin},
  {Chiaramida}, {Chiavassa}, {Chornay}, {Comoretto}, {Contursi}, {Cooper},
  {Cornez}, {Cowell}, {Crifo}, {Cropper}, {Crosta}, {Crowley}, {Dafonte},
  {Dapergolas}, {David}, {David}, {de Laverny}, {De Luise}, {De March}, {De
  Ridder}, {de Souza}, {de Torres}, {del Peloso}, {del Pozo}, {Delbo},
  {Delgado}, {Delisle}, {Demouchy}, {Dharmawardena}, {Di Matteo}, {Diakite},
  {Diener}, {Distefano}, {Dolding}, {Edvardsson}, {Enke}, {Fabre}, {Fabrizio},
  {Faigler}, {Fedorets}, {Fernique}, {Fienga}, {Figueras}, {Fournier},
  {Fouron}, {Fragkoudi}, {Gai}, {Garcia-Gutierrez}, {Garcia-Reinaldos},
  {Garc{\'\i}a-Torres}, {Garofalo}, {Gavel}, {Gavras}, {Gerlach}, {Geyer},
  {Giacobbe}, {Gilmore}, {Girona}, {Giuffrida}, {Gomel}, {Gomez},
  {Gonz{\'a}lez-N{\'u}{\~n}ez}, {Gonz{\'a}lez-Santamar{\'\i}a},
  {Gonz{\'a}lez-Vidal}, {Granvik}, {Guillout}, {Guiraud},
  {Guti{\'e}rrez-S{\'a}nchez}, {Guy}, {Hatzidimitriou}, {Hauser}, {Haywood},
  {Helmer}, {Helmi}, {Sarmiento}, {Hidalgo}, {Hilger}, {H{\l}adczuk}, {Hobbs},
  {Holland}, {Huckle}, {Jardine}, {Jasniewicz}, {Jean-Antoine Piccolo},
  {Jim{\'e}nez-Arranz}, {Jorissen}, {Juaristi Campillo}, {Julbe}, {Karbevska},
  {Kervella}, {Khanna}, {Kontizas}, {Kordopatis}, {Korn}, {K{\'o}sp{\'a}l},
  {Kostrzewa-Rutkowska}, {Kruszy{\'n}ska}, {Kun}, {Laizeau}, {Lambert},
  {Lanza}, {Lasne}, {Le Campion}, {Lebreton}, {Lebzelter}, {Leccia}, {Leclerc},
  {Lecoeur-Taibi}, {Liao}, {Licata}, {Lindstr{\o}m}, {Lister}, {Livanou},
  {Lobel}, {Lorca}, {Loup}, {Madrero Pardo}, {Magdaleno Romeo}, {Managau},
  {Mann}, {Manteiga}, {Marchant}, {Marconi}, {Marcos}, {Marcos Santos},
  {Mar{\'\i}n Pina}, {Marinoni}, {Marocco}, {Marshall}, {Martin Polo},
  {Mart{\'\i}n-Fleitas}, {Marton}, {Mary}, {Masip}, {Massari},
  {Mastrobuono-Battisti}, {Mazeh}, {McMillan}, {Messina}, {Michalik}, {Millar},
  {Mints}, {Molina}, {Molinaro}, {Moln{\'a}r}, {Monari}, {Mongui{\'o}},
  {Montegriffo}, {Montero}, {Mor}, {Mora}, {Morbidelli}, {Morel}, {Morris},
  {Muraveva}, {Murphy}, {Musella}, {Nagy}, {Noval}, {Oca{\~n}a}, {Ogden},
  {Ordenovic}, {Osinde}, {Pagani}, {Pagano}, {Palaversa}, {Palicio},
  {Pallas-Quintela}, {Panahi}, {Payne-Wardenaar}, {Pe{\~n}alosa Esteller},
  {Penttil{\"a}}, {Pichon}, {Piersimoni}, {Pineau}, {Plachy}, {Plum}, {Poggio},
  {Pr{\v{s}}a}, {Pulone}, {Racero}, {Ragaini}, {Rainer}, {Raiteri}, {Rambaux},
  {Ramos}, {Ramos-Lerate}, {Re Fiorentin}, {Regibo}, {Richards}, {Rios Diaz},
  {Ripepi}, {Riva}, {Rix}, {Rixon}, {Robichon}, {Robin}, {Robin}, {Roelens},
  {Rogues}, {Rohrbasser}, {Romero-G{\'o}mez}, {Rowell}, {Royer}, {Ruz Mieres},
  {Rybicki}, {Sadowski}, {S{\'a}ez N{\'u}{\~n}ez}, {Sagrist{\`a} Sell{\'e}s},
  {Sahlmann}, {Salguero}, {Samaras}, {Sanchez Gimenez}, {Sanna},
  {Santove{\~n}a}, {Sarasso}, {Schultheis}, {Sciacca}, {Segol}, {Segovia},
  {S{\'e}gransan}, {Semeux}, {Shahaf}, {Siddiqui}, {Siebert}, {Siltala},
  {Silvelo}, {Slezak}, {Slezak}, {Smart}, {Snaith}, {Solano}, {Solitro},
  {Souami}, {Souchay}, {Spagna}, {Spina}, {Spoto}, {Steele},
  {Steidelm{\"u}ller}, {Stephenson}, {S{\"u}veges}, {Surdej}, {Szabados},
  {Szegedi-Elek}, {Taris}, {Taylor}, {Teixeira}, {Tolomei}, {Tonello}, {Torra},
  {Torra}, {Torralba Elipe}, {Trabucchi}, {Tsounis}, {Turon}, {Ulla}, {Unger},
  {Vaillant}, {van Dillen}, {van Reeven}, {Vanel}, {Vecchiato}, {Viala},
  {Vicente}, {Voutsinas}, {Weiler}, {Wevers}, {Wyrzykowski}, {Yoldas}, {Yvard},
  {Zhao}, {Zorec}, {Zucker}, \& {Zwitter}}]{gaia2023}
{Gaia Collaboration}, {Vallenari}, A., {Brown}, A.~G.~A., {et~al.} 2023, \aap,
  674, A1

\bibitem[{{Gallenne} {et~al.}(2015){Gallenne}, {M{\'e}rand}, {Kervella},
  {Monnier}, {Schaefer}, {Baron}, {Breitfelder}, {Le Bouquin}, {Roettenbacher},
  {Gieren}, {Pietrzy{\'n}ski}, {McAlister}, {ten Brummelaar}, {Sturmann},
  {Sturmann}, {Turner}, {Ridgway}, \& {Kraus}}]{gallenne2015}
{Gallenne}, A., {M{\'e}rand}, A., {Kervella}, P., {et~al.} 2015, \aap, 579, A68

\bibitem[{{Garufi} {et~al.}(2018){Garufi}, {Benisty}, {Pinilla}, {Tazzari},
  {Dominik}, {Ginski}, {Henning}, {Kral}, {Langlois}, {M{\'e}nard}, {Stolker},
  {Szulagyi}, {Villenave}, \& {van der Plas}}]{garufi2018}
{Garufi}, A., {Benisty}, M., {Pinilla}, P., {et~al.} 2018, \aap, 620, A94

\bibitem[{{Garufi} {et~al.}(2016){Garufi}, {Quanz}, {Schmid}, {Mulders},
  {Avenhaus}, {Boccaletti}, {Ginski}, {Langlois}, {Stolker}, {Augereau},
  {Benisty}, {Lopez}, {Dominik}, {Gratton}, {Henning}, {Janson}, {M{\'e}nard},
  {Meyer}, {Pinte}, {Sissa}, {Vigan}, {Zurlo}, {Bazzon}, {Buenzli}, {Bonnefoy},
  {Brandner}, {Chauvin}, {Cheetham}, {Cudel}, {Desidera}, {Feldt}, {Galicher},
  {Kasper}, {Lagrange}, {Lannier}, {Maire}, {Mesa}, {Mouillet}, {Peretti},
  {Perrot}, {Salter}, \& {Wildi}}]{garufi2016}
{Garufi}, A., {Quanz}, S.~P., {Schmid}, H.~M., {et~al.} 2016, \aap, 588, A8

\bibitem[{{Gonzalez} {et~al.}(2020){Gonzalez}, {van der Plas}, {Pinte},
  {Cuello}, {Nealon}, {M{\'e}nard}, {Revol}, {Rodet}, {Langlois}, \&
  {Maire}}]{Gonzalez2020}
{Gonzalez}, J.-F., {van der Plas}, G., {Pinte}, C., {et~al.} 2020, \mnras, 499,
  3837

\bibitem[{{GRAVITY Collaboration} {et~al.}(2019){GRAVITY Collaboration},
  {Perraut}, {Labadie}, {Lazareff}, {Klarmann}, {Segura-Cox}, {Benisty},
  {Bouvier}, {Brandner}, {Caratti O Garatti}, {Caselli}, {Dougados}, {Garcia},
  {Garcia-Lopez}, {Kendrew}, {Koutoulaki}, {Kervella}, {Lin}, {Pineda},
  {Sanchez-Bermudez}, {van Dishoeck}, {Abuter}, {Amorim}, {Berger}, {Bonnet},
  {Buron}, {Cantalloube}, {Cl{\'e}net}, {Coud{\'e} Du Foresto}, {Dexter}, {de
  Zeeuw}, {Duvert}, {Eckart}, {Eisenhauer}, {Eupen}, {Gao}, {Gendron},
  {Genzel}, {Gillessen}, {Gordo}, {Grellmann}, {Haubois}, {Haussmann},
  {Henning}, {Hippler}, {Horrobin}, {Hubert}, {Jocou}, {Lacour}, {Le Bouquin},
  {L{\'e}na}, {M{\'e}rand}, {Ott}, {Paumard}, {Perrin}, {Pfuhl}, {Rabien},
  {Ray}, {Rau}, {Rousset}, {Scheithauer}, {Straub}, {Straubmeier}, {Sturm},
  {Vincent}, {Waisberg}, {Wank}, {Widmann}, {Wieprecht}, {Wiest}, {Wiezorrek},
  {Woillez}, \& {Yazici}}]{gravity_herbig_2019}
{GRAVITY Collaboration}, {Perraut}, K., {Labadie}, L., {et~al.} 2019, \aap,
  632, A53

\bibitem[{{Haffert} {et~al.}(2019){Haffert}, {Bohn}, {de Boer}, {Snellen},
  {Brinchmann}, {Girard}, {Keller}, \& {Bacon}}]{haffert2019}
{Haffert}, S.~Y., {Bohn}, A.~J., {de Boer}, J., {et~al.} 2019, Nature
  Astronomy, 3, 749

\bibitem[{{Hammond} {et~al.}(2022){Hammond}, {Christiaens}, {Price},
  {Ubeira-Gabellini}, {Baird}, {Calcino}, {Benisty}, {Lodato}, {Testi},
  {Pinte}, {Toci}, \& {Fedele}}]{Hammond2022}
{Hammond}, I., {Christiaens}, V., {Price}, D.~J., {et~al.} 2022, \mnras, 515,
  6109

\bibitem[{{Hinkley} {et~al.}(2011){Hinkley}, {Carpenter}, {Ireland}, \&
  {Kraus}}]{hinkley2011}
{Hinkley}, S., {Carpenter}, J.~M., {Ireland}, M.~J., \& {Kraus}, A.~L. 2011,
  \apjl, 730, L21

\bibitem[{{Hinkley} {et~al.}(2015){Hinkley}, {Kraus}, {Ireland}, {Cheetham},
  {Carpenter}, {Tuthill}, {Lacour}, {Evans}, \& {Haubois}}]{hinkley2015}
{Hinkley}, S., {Kraus}, A.~L., {Ireland}, M.~J., {et~al.} 2015, \apjl, 806, L9

\bibitem[{{Hu{\'e}lamo} {et~al.}(2011){Hu{\'e}lamo}, {Lacour}, {Tuthill},
  {Ireland}, {Kraus}, \& {Chauvin}}]{huelamo2011}
{Hu{\'e}lamo}, N., {Lacour}, S., {Tuthill}, P., {et~al.} 2011, \aap, 528, L7

\bibitem[{{Ireland}(2016)}]{ireland2016}
{Ireland}, M.~J. 2016, in Astrophysics and Space Science Library, Vol. 439,
  Astronomy at High Angular Resolution, ed. H.~M.~J. {Boffin}, G.~{Hussain},
  J.-P. {Berger}, \& L.~{Schmidtobreick}, 43

\bibitem[{{Ireland} \& {Kraus}(2008)}]{ireland2008}
{Ireland}, M.~J. \& {Kraus}, A.~L. 2008, \apjl, 678, L59

\bibitem[{{Juh{\'a}sz} \& {Facchini}(2017)}]{juhasz2017}
{Juh{\'a}sz}, A. \& {Facchini}, S. 2017, \mnras, 466, 4053

\bibitem[{{Kammerer} {et~al.}(2023){Kammerer}, {Cooper}, {Vandal}, {Thatte},
  {Martinache}, {Sivaramakrishnan}, {Chaushev}, {Stolker}, {Lloyd}, {Albert},
  {Doyon}, {Sallum}, {Perrin}, {Pueyo}, {M{\'e}rand}, {Gallenne}, {Greenbaum},
  {Sanchez-Bermudez}, {Blakely}, {Johnstone}, {Volk}, {Martel}, {Goudfrooij},
  {Meyer}, {Willott}, {De Furio}, {Dang}, {Radica}, \& {Noirot}}]{kammerer2023}
{Kammerer}, J., {Cooper}, R.~A., {Vandal}, T., {et~al.} 2023, \pasp, 135,
  014502

\bibitem[{{Kammerer} {et~al.}(2019){Kammerer}, {Ireland}, {Martinache}, \&
  {Girard}}]{kammerer2019}
{Kammerer}, J., {Ireland}, M.~J., {Martinache}, F., \& {Girard}, J.~H. 2019,
  \mnras, 486, 639

\bibitem[{{Kammerer} {et~al.}(2020){Kammerer}, {M{\'e}rand}, {Ireland}, \&
  {Lacour}}]{kammerer2020}
{Kammerer}, J., {M{\'e}rand}, A., {Ireland}, M.~J., \& {Lacour}, S. 2020, \aap,
  644, A110

\bibitem[{{Keppler} {et~al.}(2018){Keppler}, {Benisty}, {M{\"u}ller},
  {Henning}, {van Boekel}, {Cantalloube}, {Ginski}, {van Holstein}, {Maire},
  {Pohl}, {Samland }, {Avenhaus}, {Baudino}, {Boccaletti}, {de Boer},
  {Bonnefoy}, {Chauvin}, {Desidera}, {Langlois}, {Lazzoni}, {Marleau},
  {Mordasini}, {Pawellek}, {Stolker}, {Vigan}, {Zurlo}, {Birnstiel},
  {Brandner}, {Feldt}, {Flock}, {Girard}, {Gratton}, {Hagelberg}, {Isella},
  {Janson}, {Juhasz}, {Kemmer}, {Kral}, {Lagrange}, {Launhardt}, {Matter},
  {M{\'e}nard}, {Milli}, {Molli{\`e}re}, {Olofsson}, {P{\'e}rez}, {Pinilla},
  {Pinte}, {Quanz}, {Schmidt}, {Udry}, {Wahhaj}, {Williams}, {Buenzli},
  {Cudel}, {Dominik}, {Galicher}, {Kasper}, {Lannier}, {Mesa}, {Mouillet},
  {Peretti}, {Perrot}, {Salter}, {Sissa}, {Wildi}, {Abe}, {Antichi},
  {Augereau}, {Baruffolo}, {Baudoz}, {Bazzon}, {Beuzit}, {Blanchard}, {Brems},
  {Buey}, {De Caprio}, {Carbillet}, {Carle}, {Cascone}, {Cheetham}, {Claudi},
  {Costille}, {Delboulb{\'e}}, {Dohlen}, {Fantinel}, {Feautrier}, {Fusco},
  {Giro}, {Gluck}, {Gry}, {Hubin}, {Hugot}, {Jaquet}, {Le Mignant}, {Llored},
  {Madec}, {Magnard}, {Martinez}, {Maurel}, {Meyer}, {M{\"o}ller-Nilsson},
  {Moulin}, {Mugnier}, {Orign{\'e}}, {Pavlov}, {Perret}, {Petit}, {Pragt},
  {Puget}, {Rabou}, {Ramos}, {Rigal}, {Rochat}, {Roelfsema}, {Rousset}, {Roux},
  {Salasnich}, {Sauvage}, {Sevin}, {Soenke}, {Stadler}, {Suarez}, {Turatto}, \&
  {Weber}}]{keppler2018}
{Keppler}, M., {Benisty}, M., {M{\"u}ller}, A., {et~al.} 2018, \aap, 617, A44

\bibitem[{{Kraus} \& {Ireland}(2012)}]{kraus2012}
{Kraus}, A.~L. \& {Ireland}, M.~J. 2012, \apj, 745, 5

\bibitem[{{Kraus} {et~al.}(2011){Kraus}, {Ireland}, {Martinache}, \&
  {Hillenbrand}}]{kraus2011}
{Kraus}, A.~L., {Ireland}, M.~J., {Martinache}, F., \& {Hillenbrand}, L.~A.
  2011, \apj, 731, 8

\bibitem[{{Kraus} {et~al.}(2008){Kraus}, {Ireland}, {Martinache}, \&
  {Lloyd}}]{kraus2008}
{Kraus}, A.~L., {Ireland}, M.~J., {Martinache}, F., \& {Lloyd}, J.~P. 2008,
  \apj, 679, 762

\bibitem[{{Kraus} {et~al.}(2017){Kraus}, {Kreplin}, {Fukugawa}, {Muto},
  {Sitko}, {Young}, {Bate}, {Grady}, {Harries}, {Monnier}, {Willson}, \&
  {Wisniewski}}]{Kraus2017}
{Kraus}, S., {Kreplin}, A., {Fukugawa}, M., {et~al.} 2017, \apjl, 848, L11

\bibitem[{{Lacour} {et~al.}(2016){Lacour}, {Biller}, {Cheetham}, {Greenbaum},
  {Pearce}, {Marino}, {Tuthill}, {Pueyo}, {Mamajek}, {Girard},
  {Sivaramakrishnan}, {Bonnefoy}, {Baraffe}, {Chauvin}, {Olofsson}, {Juhasz},
  {Benisty}, {Pott}, {Sicilia-Aguilar}, {Henning}, {Cardwell}, {Goodsell},
  {Graham}, {Hibon}, {Ingraham}, {Konopacky}, {Macintosh}, {Oppenheimer},
  {Perrin}, {Rantakyr{\"o}}, {Sadakuni}, \& {Thomas}}]{lacour2016}
{Lacour}, S., {Biller}, B., {Cheetham}, A., {et~al.} 2016, \aap, 590, A90

\bibitem[{{Lacour} {et~al.}(2011){Lacour}, {Tuthill}, {Amico}, {Ireland},
  {Ehrenreich}, {Huelamo}, \& {Lagrange}}]{lacour2011}
{Lacour}, S., {Tuthill}, P., {Amico}, P., {et~al.} 2011, \aap, 532, A72

\bibitem[{{Maire} {et~al.}(2016){Maire}, {Langlois}, {Dohlen}, {Lagrange},
  {Gratton}, {Chauvin}, {Desidera}, {Girard}, {Milli}, {Vigan}, {Zins},
  {Delorme}, {Beuzit}, {Claudi}, {Feldt}, {Mouillet}, {Puget}, {Turatto}, \&
  {Wildi}}]{maire2016b}
{Maire}, A.-L., {Langlois}, M., {Dohlen}, K., {et~al.} 2016, in Society of
  Photo-Optical Instrumentation Engineers (SPIE) Conference Series, Vol. 9908,
  \procspie, 990834

\bibitem[{{Manara} {et~al.}(2014){Manara}, {Testi}, {Natta}, {Rosotti},
  {Benisty}, {Ercolano}, \& {Ricci}}]{Manara2014}
{Manara}, C.~F., {Testi}, L., {Natta}, A., {et~al.} 2014, \aap, 568, A18

\bibitem[{{Marino} {et~al.}(2015){Marino}, {Perez}, \& {Casassus}}]{marino2015}
{Marino}, S., {Perez}, S., \& {Casassus}, S. 2015, \apjl, 798, L44

\bibitem[{{Martinache}(2010)}]{martinache2010}
{Martinache}, F. 2010, \apj, 724, 464

\bibitem[{{Mendigut{\'\i}a} {et~al.}(2014){Mendigut{\'\i}a}, {Fairlamb},
  {Montesinos}, {Oudmaijer}, {Najita}, {Brittain}, \& {van den
  Ancker}}]{mendigutia2014}
{Mendigut{\'\i}a}, I., {Fairlamb}, J., {Montesinos}, B., {et~al.} 2014, \apj,
  790, 21

\bibitem[{{Mesa} {et~al.}(2019){Mesa}, {Keppler}, {Cantalloube}, {Rodet},
  {Charnay}, {Gratton}, {Langlois}, {Boccaletti}, {Bonnefoy}, {Vigan},
  {Flasseur}, {Bae}, {Benisty}, {Chauvin}, {de Boer}, {Desidera}, {Henning},
  {Lagrange}, {Meyer}, {Milli}, {M{\"u}ller}, {Pairet}, {Zurlo}, {Antoniucci},
  {Baudino}, {Brown Sevilla}, {Cascone}, {Cheetham}, {Claudi}, {Delorme},
  {D'Orazi}, {Feldt}, {Hagelberg}, {Janson}, {Kral}, {Lagadec}, {Lazzoni},
  {Ligi}, {Maire}, {Martinez}, {Menard}, {Meunier}, {Perrot}, {Petrus},
  {Pinte}, {Rickman}, {Rochat}, {Rouan}, {Samland}, {Sauvage}, {Schmidt},
  {Udry}, {Weber}, \& {Wildi}}]{mesa2019}
{Mesa}, D., {Keppler}, M., {Cantalloube}, F., {et~al.} 2019, \aap, 632, A25

\bibitem[{{Min} {et~al.}(2017){Min}, {Stolker}, {Dominik}, \&
  {Benisty}}]{min2017}
{Min}, M., {Stolker}, T., {Dominik}, C., \& {Benisty}, M. 2017, \aap, 604, L10

\bibitem[{{M{\"u}ller} {et~al.}(2018){M{\"u}ller}, {Keppler}, {Henning},
  {Samland}, {Chauvin}, {Beust}, {Maire}, {Molaverdikhani}, {van Boekel},
  {Benisty}, {Boccaletti}, {Bonnefoy}, {Cantalloube}, {Charnay}, {Baudino},
  {Gennaro}, {Long}, {Cheetham}, {Desidera}, {Feldt}, {Fusco}, {Girard},
  {Gratton}, {Hagelberg}, {Janson}, {Lagrange}, {Langlois}, {Lazzoni}, {Ligi},
  {M{\'e}nard}, {Mesa}, {Meyer}, {Molli{\`e}re}, {Mordasini}, {Moulin},
  {Pavlov}, {Pawellek}, {Quanz}, {Ramos}, {Rouan}, {Sissa}, {Stadler}, {Vigan},
  {Wahhaj}, {Weber}, \& {Zurlo}}]{mueller2018}
{M{\"u}ller}, A., {Keppler}, M., {Henning}, T., {et~al.} 2018, \aap, 617, L2

\bibitem[{{Olofsson} {et~al.}(2013){Olofsson}, {Benisty}, {Le Bouquin},
  {Berger}, {Lacour}, {M{\'e}nard}, {Henning}, {Crida}, {Burtscher}, {Meeus},
  {Ratzka}, {Pinte}, {Augereau}, {Malbet}, {Lazareff}, \&
  {Traub}}]{olofsson2013}
{Olofsson}, J., {Benisty}, M., {Le Bouquin}, J.~B., {et~al.} 2013, \aap, 552,
  A4

\bibitem[{{Perez} {et~al.}(2015){Perez}, {Casassus}, {M{\'e}nard}, {Roman},
  {van der Plas}, {Cieza}, {Pinte}, {Christiaens}, \& {Hales}}]{perez2015}
{Perez}, S., {Casassus}, S., {M{\'e}nard}, F., {et~al.} 2015, \apj, 798, 85

\bibitem[{{Pinilla} {et~al.}(2022){Pinilla}, {Benisty}, {Kurtovic}, {Bae},
  {Dong}, {Zhu}, {Andrews}, {Carpenter}, {Ginski}, {Huang}, {Isella},
  {P{\'e}rez}, {Ricci}, {Rosotti}, {Villenave}, \& {Wilner}}]{Pinilla2022}
{Pinilla}, P., {Benisty}, M., {Kurtovic}, N.~T., {et~al.} 2022, \aap, 665, A128

\bibitem[{{Price} {et~al.}(2018){Price}, {Cuello}, {Pinte}, {Mentiplay},
  {Casassus}, {Christiaens}, {Kennedy}, {Cuadra}, {Sebastian Perez}, {Marino},
  {Armitage}, {Zurlo}, {Juhasz}, {Ragusa}, {Laibe}, \& {Lodato}}]{price2018}
{Price}, D.~J., {Cuello}, N., {Pinte}, C., {et~al.} 2018, \mnras, 477, 1270

\bibitem[{{Rabago} {et~al.}(2023){Rabago}, {Zhu}, {Martin}, \&
  {Lubow}}]{Rabago2023}
{Rabago}, I., {Zhu}, Z., {Martin}, R.~G., \& {Lubow}, S.~H. 2023, \mnras, 520,
  2138

\bibitem[{{Ragusa} {et~al.}(2020){Ragusa}, {Alexander}, {Calcino}, {Hirsh}, \&
  {Price}}]{ragusa2020}
{Ragusa}, E., {Alexander}, R., {Calcino}, J., {Hirsh}, K., \& {Price}, D.~J.
  2020, \mnras, 499, 3362

\bibitem[{{Ragusa} {et~al.}(2017){Ragusa}, {Dipierro}, {Lodato}, {Laibe}, \&
  {Price}}]{ragusa2017}
{Ragusa}, E., {Dipierro}, G., {Lodato}, G., {Laibe}, G., \& {Price}, D.~J.
  2017, \mnras, 464, 1449

\bibitem[{{Rayner} {et~al.}(2003){Rayner}, {Toomey}, {Onaka}, {Denault},
  {Stahlberger}, {Vacca}, {Cushing}, \& {Wang}}]{rayner2003}
{Rayner}, J.~T., {Toomey}, D.~W., {Onaka}, P.~M., {et~al.} 2003, \pasp, 115,
  362

\bibitem[{{Rich} {et~al.}(2022){Rich}, {Monnier}, {Aarnio}, {Laws},
  {Setterholm}, {Wilner}, {Calvet}, {Harries}, {Miller}, {Davies}, {Adams},
  {Andrews}, {Bae}, {Espaillat}, {Greenbaum}, {Hinkley}, {Kraus}, {Hartmann},
  {Isella}, {McClure}, {Oppenheimer}, {P{\'e}rez}, \& {Zhu}}]{rich2022}
{Rich}, E.~A., {Monnier}, J.~D., {Aarnio}, A., {et~al.} 2022, \aj, 164, 109

\bibitem[{{Riols} \& {Lesur}(2019)}]{riols2019}
{Riols}, A. \& {Lesur}, G. 2019, \aap, 625, A108

\bibitem[{{Rodigas} {et~al.}(2014){Rodigas}, {Follette}, {Weinberger}, {Close},
  \& {Hines}}]{rodigas2014}
{Rodigas}, T.~J., {Follette}, K.~B., {Weinberger}, A., {Close}, L., \& {Hines},
  D.~C. 2014, \apjl, 791, L37

\bibitem[{{Rosotti} {et~al.}(2020){Rosotti}, {Benisty}, {Juh{\'a}sz}, {Teague},
  {Clarke}, {Dominik}, {Dullemond}, {Klaassen}, {Matr{\`a}}, \&
  {Stolker}}]{Rosotti2020}
{Rosotti}, G.~P., {Benisty}, M., {Juh{\'a}sz}, A., {et~al.} 2020, \mnras, 491,
  1335

\bibitem[{{Ru{\'\i}z-Rodr{\'\i}guez} {et~al.}(2016){Ru{\'\i}z-Rodr{\'\i}guez},
  {Ireland}, {Cieza}, \& {Kraus}}]{ruiz2016}
{Ru{\'\i}z-Rodr{\'\i}guez}, D., {Ireland}, M., {Cieza}, L., \& {Kraus}, A.
  2016, \mnras, 463, 3829

\bibitem[{{Sallum} {et~al.}(2023){Sallum}, {Eisner}, {Skemer}, \&
  {Murray-Clay}}]{sallum2023}
{Sallum}, S., {Eisner}, J., {Skemer}, A., \& {Murray-Clay}, R. 2023, \apj, 953,
  55

\bibitem[{{Sallum} {et~al.}(2015{\natexlab{a}}){Sallum}, {Eisner}, {Close},
  {Hinz}, {Skemer}, {Bailey}, {Briguglio}, {Follette}, {Males}, {Morzinski},
  {Puglisi}, {Rodigas}, {Weinberger}, \& {Xompero}}]{sallum2015a}
{Sallum}, S., {Eisner}, J.~A., {Close}, L.~M., {et~al.} 2015{\natexlab{a}},
  \apj, 801, 85

\bibitem[{{Sallum} {et~al.}(2015{\natexlab{b}}){Sallum}, {Follette}, {Eisner},
  {Close}, {Hinz}, {Kratter}, {Males}, {Skemer}, {Macintosh}, {Tuthill},
  {Bailey}, {Defr{\`e}re}, {Morzinski}, {Rodigas}, {Spalding}, {Vaz}, \&
  {Weinberger}}]{sallum2015b}
{Sallum}, S., {Follette}, K.~B., {Eisner}, J.~A., {et~al.} 2015{\natexlab{b}},
  \nat, 527, 342

\bibitem[{{Sanchis} {et~al.}(2020){Sanchis}, {Picogna}, {Ercolano}, {Testi}, \&
  {Rosotti}}]{sanchis2020}
{Sanchis}, E., {Picogna}, G., {Ercolano}, B., {Testi}, L., \& {Rosotti}, G.
  2020, \mnras, 492, 3440

\bibitem[{{Sivaramakrishnan} {et~al.}(2023){Sivaramakrishnan}, {Tuthill},
  {Lloyd}, {Greenbaum}, {Thatte}, {Cooper}, {Vandal}, {Kammerer},
  {Sanchez-Bermudez}, {Pope}, {Blakely}, {Albert}, {Cook}, {Johnstone},
  {Martel}, {Volk}, {Soulain}, {Artigau}, {Lafreni{\`e}re}, {Willott},
  {Parmentier}, {Ford}, {McKernan}, {Vila}, {Rowlands}, {Doyon}, {Beaulieu},
  {Desdoigts}, {Fullerton}, {De Furio}, {Goudfrooij}, {Holfeltz}, {LaMassa},
  {Maszkiewicz}, {Meyer}, {Perrin}, {Pueyo}, {Sahlmann}, {Sohn}, {Teixeira}, \&
  {Zheng}}]{sivaramakrishnan2023}
{Sivaramakrishnan}, A., {Tuthill}, P., {Lloyd}, J.~P., {et~al.} 2023, \pasp,
  135, 015003

\bibitem[{{Skrutskie} {et~al.}(1990){Skrutskie}, {Dutkevitch}, {Strom},
  {Edwards}, {Strom}, \& {Shure}}]{skrutskie1990}
{Skrutskie}, M.~F., {Dutkevitch}, D., {Strom}, S.~E., {et~al.} 1990, \aj, 99,
  1187

\bibitem[{{Soulain} {et~al.}(2020){Soulain}, {Sivaramakrishnan}, {Tuthill},
  {Thatte}, {Volk}, {Cooper}, {Albert}, {Artigau}, {Cook}, {Doyon},
  {Johnstone}, {Lafreni{\`e}re}, \& {Martel}}]{soulain2020}
{Soulain}, A., {Sivaramakrishnan}, A., {Tuthill}, P., {et~al.} 2020, in Society
  of Photo-Optical Instrumentation Engineers (SPIE) Conference Series, Vol.
  11446, Society of Photo-Optical Instrumentation Engineers (SPIE) Conference
  Series, 1144611

\bibitem[{{Stolker} {et~al.}(2016){Stolker}, {Dominik}, {Avenhaus}, {Min}, {de
  Boer}, {Ginski}, {Schmid}, {Juhasz}, {Bazzon}, {Waters}, {Garufi},
  {Augereau}, {Benisty}, {Boccaletti}, {Henning}, {Langlois}, {Maire},
  {M{\'e}nard}, {Meyer}, {Pinte}, {Quanz}, {Thalmann}, {Beuzit}, {Carbillet},
  {Costille}, {Dohlen}, {Feldt}, {Gisler}, {Mouillet}, {Pavlov}, {Perret},
  {Petit}, {Pragt}, {Rochat}, {Roelfsema}, {Salasnich}, {Soenke}, \&
  {Wildi}}]{stolker2016b}
{Stolker}, T., {Dominik}, C., {Avenhaus}, H., {et~al.} 2016, \aap, 595, A113

\bibitem[{{Stolker} {et~al.}(2020{\natexlab{a}}){Stolker}, {Marleau}, {Cugno},
  {Molli{\`e}re}, {Quanz}, {Todorov}, \& {K{\"u}hn}}]{stolker2020b}
{Stolker}, T., {Marleau}, G.-D., {Cugno}, G., {et~al.} 2020{\natexlab{a}},
  \aap, 644, A13

\bibitem[{{Stolker} {et~al.}(2020{\natexlab{b}}){Stolker}, {Quanz}, {Todorov},
  {K{\"u}hn}, {Molli{\`e}re}, {Meyer}, {Currie}, {Daemgen}, \&
  {Lavie}}]{stolker2020a}
{Stolker}, T., {Quanz}, S.~P., {Todorov}, K.~O., {et~al.} 2020{\natexlab{b}},
  \aap, 635, A182

\bibitem[{{Stolker} {et~al.}(2017){Stolker}, {Sitko}, {Lazareff}, {Benisty},
  {Dominik}, {Waters}, {Min}, {Perez}, {Milli}, {Garufi}, {de Boer}, {Ginski},
  {Kraus}, {Berger}, \& {Avenhaus}}]{stolker2017}
{Stolker}, T., {Sitko}, M., {Lazareff}, B., {et~al.} 2017, \apj, 849, 143

\bibitem[{{Strom} {et~al.}(1989){Strom}, {Strom}, {Edwards}, {Cabrit}, \&
  {Skrutskie}}]{strom1989}
{Strom}, K.~M., {Strom}, S.~E., {Edwards}, S., {Cabrit}, S., \& {Skrutskie},
  M.~F. 1989, \aj, 97, 1451

\bibitem[{{Toci} {et~al.}(2020){Toci}, {Lodato}, {Christiaens}, {Fedele},
  {Pinte}, {Price}, \& {Testi}}]{Toci2020}
{Toci}, C., {Lodato}, G., {Christiaens}, V., {et~al.} 2020, \mnras, 499, 2015

\bibitem[{{Trotta}(2008)}]{trotta2008}
{Trotta}, R. 2008, Contemporary Physics, 49, 71

\bibitem[{{Vacca} {et~al.}(2003){Vacca}, {Cushing}, \& {Rayner}}]{vacca2003}
{Vacca}, W.~D., {Cushing}, M.~C., \& {Rayner}, J.~T. 2003, \pasp, 115, 389

\bibitem[{{van der Marel} {et~al.}(2016){van der Marel}, {van Dishoeck},
  {Bruderer}, {Andrews}, {Pontoppidan}, {Herczeg}, {van Kempen}, \&
  {Miotello}}]{vandermarel2016}
{van der Marel}, N., {van Dishoeck}, E.~F., {Bruderer}, S., {et~al.} 2016,
  \aap, 585, A58

\bibitem[{{van Holstein} {et~al.}(2021){van Holstein}, {Stolker},
  {Jensen-Clem}, {Ginski}, {Milli}, {de Boer}, {Girard}, {Wahhaj}, {Bohn},
  {Millar-Blanchaer}, {Benisty}, {Bonnefoy}, {Chauvin}, {Dominik}, {Hinkley},
  {Keller}, {Keppler}, {Langlois}, {Marino}, {M{\'e}nard}, {Perrot}, {Schmidt},
  {Vigan}, {Zurlo}, \& {Snik}}]{vanholstein2021}
{van Holstein}, R.~G., {Stolker}, T., {Jensen-Clem}, R., {et~al.} 2021, \aap,
  647, A21

\bibitem[{{Verhoeff} {et~al.}(2011){Verhoeff}, {Min}, {Pantin}, {Waters},
  {Tielens}, {Honda}, {Fujiwara}, {Bouwman}, {van Boekel}, {Dougherty}, {de
  Koter}, {Dominik}, \& {Mulders}}]{verhoeff2011}
{Verhoeff}, A.~P., {Min}, M., {Pantin}, E., {et~al.} 2011, \aap, 528, A91

\bibitem[{{Vigan}(2020)}]{vigan2020}
{Vigan}, A. 2020, {vlt-sphere: Automatic VLT/SPHERE data reduction and
  analysis}, Astrophysics Source Code Library, record ascl:2009.002

\bibitem[{{Vioque} {et~al.}(2018){Vioque}, {Oudmaijer}, {Baines},
  {Mendigut{\'\i}a}, \& {P{\'e}rez-Mart{\'\i}nez}}]{vioque2018}
{Vioque}, M., {Oudmaijer}, R.~D., {Baines}, D., {Mendigut{\'\i}a}, I., \&
  {P{\'e}rez-Mart{\'\i}nez}, R. 2018, \aap, 620, A128

\bibitem[{{Vousden} {et~al.}(2016){Vousden}, {Farr}, \& {Mandel}}]{vousden2016}
{Vousden}, W.~D., {Farr}, W.~M., \& {Mandel}, I. 2016, \mnras, 455, 1919

\bibitem[{{Wahhaj} {et~al.}(2021){Wahhaj}, {Milli}, {Romero}, {Cieza}, {Zurlo},
  {Vigan}, {Pe{\~n}a}, {Valdes}, {Cantalloube}, {Girard}, \&
  {Pantoja}}]{wahhaj2021}
{Wahhaj}, Z., {Milli}, J., {Romero}, C., {et~al.} 2021, \aap, 648, A26

\bibitem[{{Wang} {et~al.}(2021){Wang}, {Vigan}, {Lacour}, {Nowak}, {Stolker},
  {De Rosa}, {Ginzburg}, {Gao}, {Abuter}, {Amorim}, {Asensio-Torres},
  {Baub{\"o}ck}, {Benisty}, {Berger}, {Beust}, {Beuzit}, {Blunt}, {Boccaletti},
  {Bohn}, {Bonnefoy}, {Bonnet}, {Brandner}, {Cantalloube}, {Caselli},
  {Charnay}, {Chauvin}, {Choquet}, {Christiaens}, {Cl{\'e}net}, {Coud{\'e} Du
  Foresto}, {Cridland}, {de Zeeuw}, {Dembet}, {Dexter}, {Drescher}, {Duvert},
  {Eckart}, {Eisenhauer}, {Facchini}, {Gao}, {Garcia}, {Garcia Lopez},
  {Gardner}, {Gendron}, {Genzel}, {Gillessen}, {Girard}, {Haubois},
  {Hei{\ss}el}, {Henning}, {Hinkley}, {Hippler}, {Horrobin}, {Houll{\'e}},
  {Hubert}, {Jim{\'e}nez-Rosales}, {Jocou}, {Kammerer}, {Keppler}, {Kervella},
  {Meyer}, {Kreidberg}, {Lagrange}, {Lapeyr{\`e}re}, {Le Bouquin}, {L{\'e}na},
  {Lutz}, {Maire}, {M{\'e}nard}, {M{\'e}rand}, {Molli{\`e}re}, {Monnier},
  {Mouillet}, {M{\"u}ller}, {Nasedkin}, {Ott}, {Otten}, {Paladini}, {Paumard},
  {Perraut}, {Perrin}, {Pfuhl}, {Pueyo}, {Rameau}, {Rodet},
  {Rodr{\'\i}guez-Coira}, {Rousset}, {Scheithauer}, {Shangguan}, {Shimizu},
  {Stadler}, {Straub}, {Straubmeier}, {Sturm}, {Tacconi}, {van Dishoeck},
  {Vincent}, {von Fellenberg}, {Ward-Duong}, {Widmann}, {Wieprecht},
  {Wiezorrek}, {Woillez}, \& {Gravity Collaboration}}]{wang2021}
{Wang}, J.~J., {Vigan}, A., {Lacour}, S., {et~al.} 2021, \aj, 161, 148

\bibitem[{{Willson} {et~al.}(2019){Willson}, {Kraus}, {Kluska}, {Monnier},
  {Cure}, {Sitko}, {Aarnio}, {Ireland}, {Rizzuto}, {Hone}, {Kreplin},
  {Andrews}, {Calvet}, {Espaillat}, {Fukagawa}, {Harries}, {Hinkley}, {Kanaan},
  {Muto}, \& {Wilner}}]{willson2019}
{Willson}, M., {Kraus}, S., {Kluska}, J., {et~al.} 2019, \aap, 621, A7

\bibitem[{{Willson} {et~al.}(2016){Willson}, {Kraus}, {Kluska}, {Monnier},
  {Ireland}, {Aarnio}, {Sitko}, {Calvet}, {Espaillat}, \&
  {Wilner}}]{willson2016}
{Willson}, M., {Kraus}, S., {Kluska}, J., {et~al.} 2016, \aap, 595, A9

\bibitem[{{Zhang} {et~al.}(2018){Zhang}, {Zhu}, {Huang}, {Guzm{\'a}n},
  {Andrews}, {Birnstiel}, {Dullemond}, {Carpenter}, {Isella}, {P{\'e}rez},
  {Benisty}, {Wilner}, {Baruteau}, {Bai}, \& {Ricci}}]{zhang2018}
{Zhang}, S., {Zhu}, Z., {Huang}, J., {et~al.} 2018, \apjl, 869, L47

\bibitem[{{Zhu}(2019)}]{Zhu2019}
{Zhu}, Z. 2019, \mnras, 483, 4221

\bibitem[{{Zhu} {et~al.}(2011){Zhu}, {Nelson}, {Hartmann}, {Espaillat}, \&
  {Calvet}}]{Zhu2011}
{Zhu}, Z., {Nelson}, R.~P., {Hartmann}, L., {Espaillat}, C., \& {Calvet}, N.
  2011, \apj, 729, 47

\end{thebibliography}

\begin{appendix}

\section{SPHERE/IFS spectrum}
\label{sec:appendix_spectrum}

Table~\ref{table:ifs_spectrum} lists the extracted contrast and derived fluxes for the SPHERE/IFS dataset from 2019\,May\,18 that is used for the spectral analysis in Sect.~\ref{sec:hd142527b_atmosphere}.

\begin{table}
\caption{Spectrophotometry of HD\,142527\,B.}
\label{table:ifs_spectrum}
\centering
\bgroup
\def\arraystretch{1.25}
\begin{tabular}{C{1.7cm} C{1.8cm} C{3cm}}
\hline\hline
Wavelength & Contrast & Flux\\
($\mu$m) & (mag) & (W~m$^{-2}$~$\mu$m$^{-1}$)\\
\hline
$0.953$ & $4.54 \pm 0.08$ & $1.77 \pm 0.13$ $\times$ $10^{-13}$ \\
$0.972$ & $4.62 \pm 0.06$ & $1.63 \pm 0.10$ $\times$ $10^{-13}$ \\
$0.991$ & $4.55 \pm 0.05$ & $1.68 \pm 0.08$ $\times$ $10^{-13}$ \\
$1.010$ & $4.46 \pm 0.04$ & $1.76 \pm 0.06$ $\times$ $10^{-13}$ \\
$1.029$ & $4.45 \pm 0.04$ & $1.76 \pm 0.06$ $\times$ $10^{-13}$ \\
$1.048$ & $4.44 \pm 0.04$ & $1.72 \pm 0.06$ $\times$ $10^{-13}$ \\
$1.067$ & $4.32 \pm 0.03$ & $1.89 \pm 0.06$ $\times$ $10^{-13}$ \\
$1.086$ & $4.39 \pm 0.04$ & $1.72 \pm 0.06$ $\times$ $10^{-13}$ \\
$1.105$ & $4.29 \pm 0.03$ & $1.83 \pm 0.05$ $\times$ $10^{-13}$ \\
$1.124$ & $4.31 \pm 0.03$ & $1.76 \pm 0.05$ $\times$ $10^{-13}$ \\
$1.144$ & $4.30 \pm 0.03$ & $1.75 \pm 0.05$ $\times$ $10^{-13}$ \\
$1.163$ & $4.22 \pm 0.02$ & $1.84 \pm 0.04$ $\times$ $10^{-13}$ \\
$1.182$ & $4.25 \pm 0.03$ & $1.75 \pm 0.04$ $\times$ $10^{-13}$ \\
$1.201$ & $4.30 \pm 0.02$ & $1.63 \pm 0.03$ $\times$ $10^{-13}$ \\
$1.220$ & $4.26 \pm 0.03$ & $1.66 \pm 0.04$ $\times$ $10^{-13}$ \\
$1.239$ & $4.31 \pm 0.02$ & $1.56 \pm 0.03$ $\times$ $10^{-13}$ \\
$1.258$ & $4.26 \pm 0.02$ & $1.61 \pm 0.03$ $\times$ $10^{-13}$ \\
$1.277$ & $4.26 \pm 0.02$ & $1.58 \pm 0.03$ $\times$ $10^{-13}$ \\
$1.296$ & $4.20 \pm 0.02$ & $1.63 \pm 0.02$ $\times$ $10^{-13}$ \\
$1.315$ & $4.28 \pm 0.02$ & $1.47 \pm 0.03$ $\times$ $10^{-13}$ \\
$1.334$ & $4.25 \pm 0.02$ & $1.48 \pm 0.03$ $\times$ $10^{-13}$ \\
$1.353$ & $4.30 \pm 0.04$ & $1.40 \pm 0.05$ $\times$ $10^{-13}$ \\
$1.372$ & $4.21 \pm 0.05$ & $1.50 \pm 0.08$ $\times$ $10^{-13}$ \\
$1.391$ & $4.28 \pm 0.07$ & $1.39 \pm 0.08$ $\times$ $10^{-13}$ \\
$1.410$ & $4.34 \pm 0.04$ & $1.30 \pm 0.04$ $\times$ $10^{-13}$ \\
$1.429$ & $4.31 \pm 0.03$ & $1.32 \pm 0.03$ $\times$ $10^{-13}$ \\
$1.448$ & $4.39 \pm 0.02$ & $1.21 \pm 0.03$ $\times$ $10^{-13}$ \\
$1.467$ & $4.33 \pm 0.02$ & $1.25 \pm 0.02$ $\times$ $10^{-13}$ \\
$1.486$ & $4.31 \pm 0.02$ & $1.27 \pm 0.02$ $\times$ $10^{-13}$ \\
$1.506$ & $4.38 \pm 0.02$ & $1.17 \pm 0.02$ $\times$ $10^{-13}$ \\
$1.525$ & $4.30 \pm 0.01$ & $1.24 \pm 0.02$ $\times$ $10^{-13}$ \\
$1.544$ & $4.29 \pm 0.02$ & $1.23 \pm 0.02$ $\times$ $10^{-13}$ \\
$1.563$ & $4.27 \pm 0.01$ & $1.23 \pm 0.02$ $\times$ $10^{-13}$ \\
$1.582$ & $4.19 \pm 0.01$ & $1.30 \pm 0.02$ $\times$ $10^{-13}$ \\
$1.601$ & $4.22 \pm 0.01$ & $1.26 \pm 0.02$ $\times$ $10^{-13}$ \\
$1.620$ & $4.18 \pm 0.02$ & $1.30 \pm 0.02$ $\times$ $10^{-13}$ \\
$1.639$ & $4.17 \pm 0.01$ & $1.31 \pm 0.02$ $\times$ $10^{-13}$ \\
$1.658$ & $4.15 \pm 0.02$ & $1.33 \pm 0.02$ $\times$ $10^{-13}$ \\
$1.677$ & $4.03 \pm 0.02$ & $1.46 \pm 0.03$ $\times$ $10^{-13}$ \\
\hline
\end{tabular}
\egroup
\end{table}

\section{Posterior distributions}
\label{sec:appendix_posteriors}

\begin{figure*}
\centering
\includegraphics[width=0.7\linewidth]{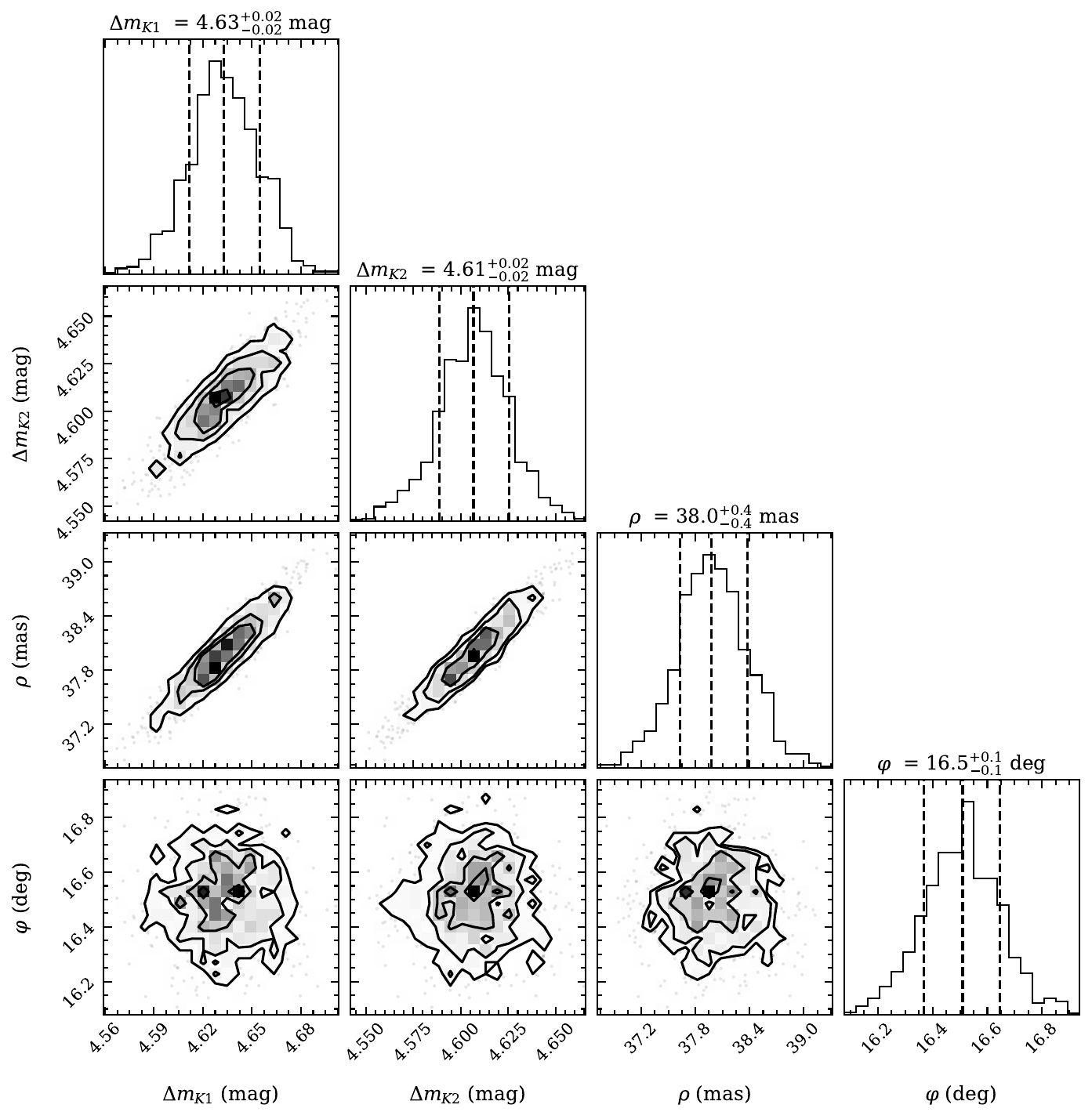}
\caption{Posterior distribution from the simultaneous extraction of the relative brightness of HD\,142527\,B in the $K12$ dual-band filters, $\Delta m_{K1}$ and $\Delta m_{K2}$, of the 2019 dataset, as well as the separation, $\rho$, and position angle, $\varphi$.}
\label{fig:posterior_extraction}
\end{figure*}

\begin{figure*}
\centering
\includegraphics[width=\linewidth]{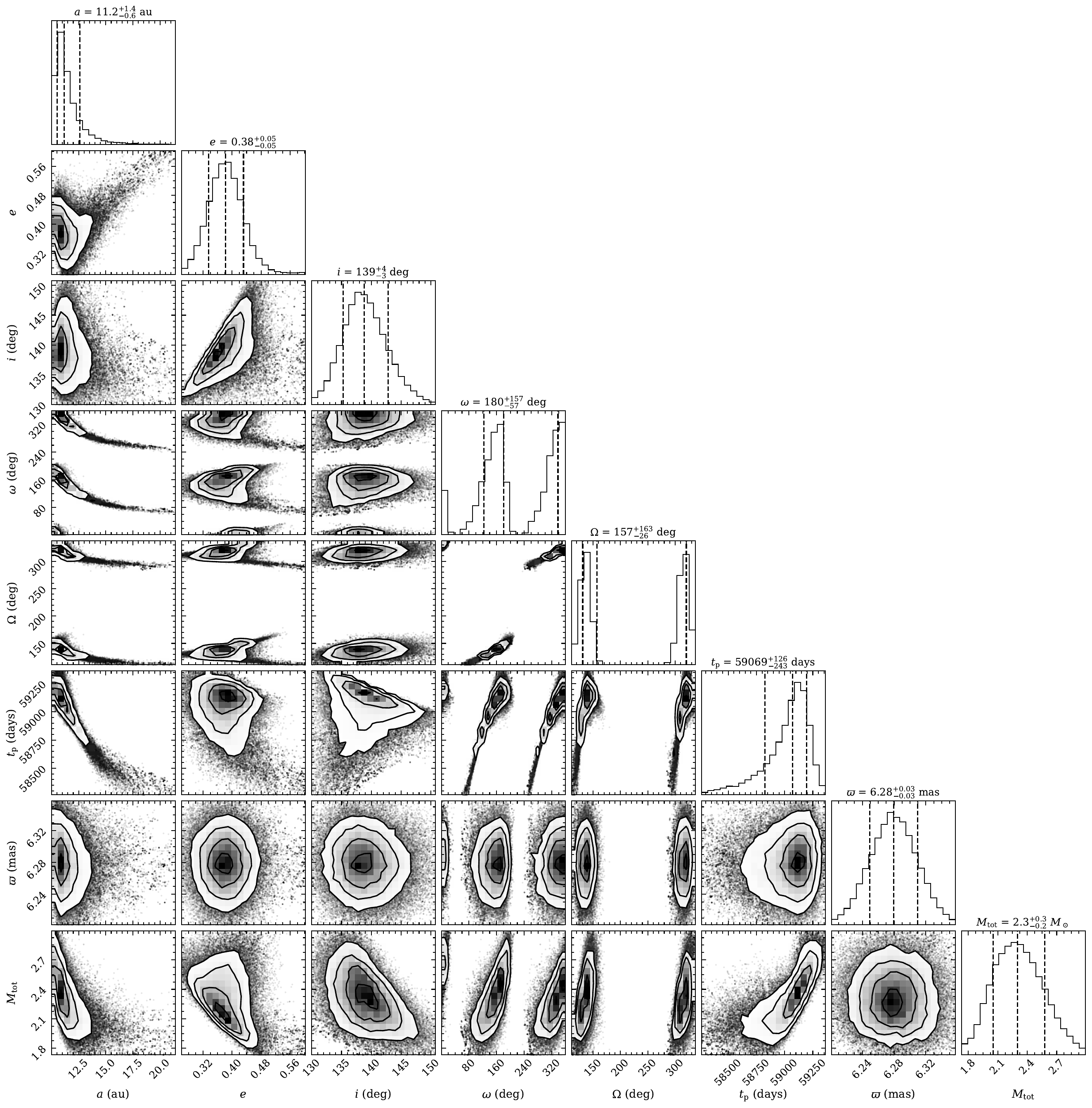}
\caption{Posterior distribution from fitting orbits to the relative astrometry of HD\,142527\,B.}
\label{fig:posterior_orbit}
\end{figure*}

\begin{figure*}
\centering
\includegraphics[width=\linewidth]{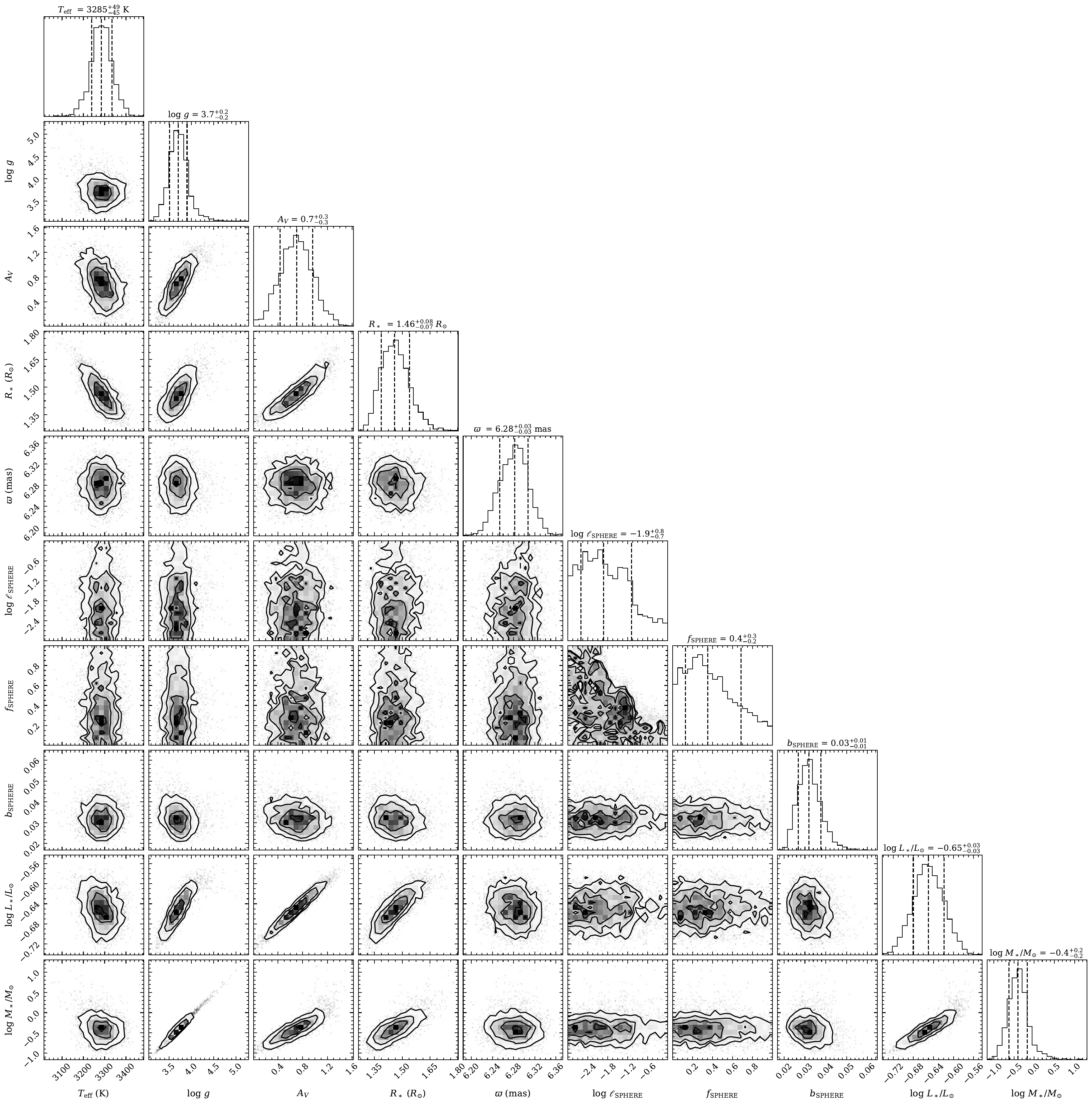}
\caption{Posterior distribution from fitting the spectral energy distribution of HD\,142527\,B with a grid of model spectra. The bolometric luminosity, $L_\ast$, and companion mass, $M_\ast$, are not free parameters with the fit but were derived afterwards from the posterior samples.}
\label{fig:posterior_sed}
\end{figure*}

In this appendix, we show in Fig.~\ref{fig:posterior_extraction}, Fig.~\ref{fig:posterior_orbit}, and Fig.~\ref{fig:posterior_sed} the 1D and 2D marginalized posterior distributions from the extracted binary parameters for one of the IRDIS datasets (see Sect.~\ref{sec:hd142527_astrometry_photometry}), the orbital analysis (see Sect.~\ref{sec:hd142527b_orbit}), and spectral analysis (see Sect.~\ref{sec:hd142527b_atmosphere}), respectively.

\end{appendix}

\end{document}